\documentclass[journal]{IEEEtran}

\usepackage{cite}
\usepackage{amsmath,amssymb,amsfonts,bbm}
\usepackage{graphicx}
\usepackage{subfigure}
\usepackage[ruled]{algorithm2e}
\usepackage{algorithmic}
\usepackage{color}

\newtheorem{corollary}{\textbf{\textit{Corollary}}}

\allowdisplaybreaks[3] % 允许长公式跨页

\usepackage[colorlinks,linkcolor=blue]{hyperref}
\normalsize

\ifCLASSINFOpdf

\else

\fi

\begin{document}
\title{Incremental Collaborative Beam Alignment for Millimeter Wave Cell-Free MIMO Systems}

\author{Cheng~Zhang,~\IEEEmembership{Member,~IEEE},
 Leming Chen, Lujia Zhang,\\ 
 Yongming Huang,~\IEEEmembership{Senior Member,~IEEE},
		Wei Zhang,~\IEEEmembership{Fellow, IEEE}
        % <-this % stops a space
\thanks{This work was supported in part by the National Key R\&D Program of China under Grant 2018YFB1800801, the National Natural Science Foundation of China under Grant 62271140 and 62225107, and the Fundamental Research Funds for the Central Universities 2242022k60002. (Corresponding authors: Y. Huang, C. Zhang)}
\thanks{C. Zhang, L. Chen, L. Zhang, and Y. Huang are with the National Mobile Communication Research Laboratory, the School of Information Science and Engineering, Southeast University, Nanjing 210096, China, and also with the Purple Mountain Laboratories, Nanjing 211111, China (e-mail: $ \left\lbrace \right.  $zhangcheng\_seu; chenleming; zhanglujia; huangym$ \left.  \right\rbrace $@seu.edu.cn).
		
W. Zhang is with the School of Electrical Engineering and Telecommunications, the University of New South Wales, Sydney, NSW 2052, Australia, and also with the Purple Mountain Laboratories, Nanjing 211111, China (e-mail: wzhang@ee.unsw.edu.au).
}}

\markboth{IEEE Transactions on Communications}%
{Submitted paper}
\maketitle

\begin{abstract}

Millimeter wave (mmWave) cell-free MIMO achieves an extremely high rate while its beam alignment (BA) suffers from excessive overhead due to a large number of transceivers. 
Recently, {user location and probing measurements} are utilized for BA based on machine learning (ML) models, e.g., deep neural network (DNN). 
However, most of these ML models are centralized with high communication and computational overhead and give no specific consideration to practical issues, e.g., limited training data and real-time model updates. 
In this paper, we study the {probing} beam-based BA for mmWave cell-free MIMO downlink with the help of broad learning (BL).
For channels without and with uplink-downlink reciprocity, we propose the user-side and base station (BS)-side BL-aided incremental collaborative BA approaches. 
Via transforming the centralized BL into a distributed learning with data and feature splitting respectively, the user-side and BS-side schemes realize implicit sharing of multiple user data and multiple BS features. 
Simulations confirm that the user-side scheme is applicable to fast time-varying and/or non-stationary channels, while the BS-side scheme is suitable for systems with low-bandwidth fronthaul links and a central unit with limited computing power.
The advantages of proposed schemes are also demonstrated compared to traditional and DNN-aided BA schemes.

\end{abstract}


% \fi
\IEEEpeerreviewmaketitle

\section{Introduction}

\IEEEPARstart{B}y exploiting the large bandwidth in the millimeter wave (mmWave) band $30-300$ GHz \cite{xiao2017millimeter}, mmWave communications can achieve multiple Gbps data rates.
 To compensate for the large attenuation and blockage effect of the mmWave signal propagation, massive multi-input-multi-output (MIMO) \cite{lu2014overview} and cell-free networks\cite{ngo2017cell} are successively introduced to provide high beamforming (BF) gain and macro-diversity \cite{alonzo2019energy}. 
 Researchers have studied several key enabling technologies, e.g., channel estimation\cite{jin2019channel}, hybrid analog-digital BF \cite{nguyen2022hybrid,yetis2021joint}, power control \cite{alonzo2019energy}, pilot allocation \cite{femenias2019cell} and user association \cite{wang2021millimeter}, for mmWave cell-free MIMO systems.
%  Some key enabling technologies, e.g., channel estimation\cite{jin2019channel}, hybrid analog-digital BF \cite{nguyen2022hybrid,yetis2021joint}, power control \cite{alonzo2019energy}, pilot allocation \cite{femenias2019cell} and user association \cite{wang2021millimeter}, have been studied for mmWave cell-free MIMO systems.

%In mmWave communications, transceivers, e.g., base stations (BSs) and user equipments (UEs), typically adopt codebooks of indexed analog beams. 
Transceivers in mmWave communications, such as base stations (BSs) and user equipments (UEs), usually utilize codebooks consisting of indexed analog beams.
The process of searching and maintaining near-optimal analog BF weights is commonly known as beam alignment (BA) \cite{wang2009beam}.
%The procedure of searching and maintaining near-optimal analog BF weights is often referred to as beam alignment (BA) \cite{wang2009beam}. 
%The most typical BA framework is based on beam sweeping, measurements, and reporting \cite{giordani2018tutorial}.
The typical BA framework involves beam sweeping, measurements, and reporting \cite{giordani2018tutorial}.
When using exhaustive beam sweeping, transceivers are required to search through all possible combinations of beam pairs.
%With exhaustive beam sweeping, transceivers need to search all combinations of beam pairs. 
Compared to exhaustive beam sweeping, hierarchical beam sweeping can reduce the overhead and latency, where beams with decreasing width are iteratively trained to identify the optimal narrow beam \cite{noh2017multi}.
%Hierarchical beam sweeping can help reduce the overhead and latency of the exhaustive one \cite{noh2017multi}, where beams with decreasing width are iteratively trained to find the optimal narrow beam. 
However, this process is susceptible to signal-to-noise ratio (SNR) during beam sweeping and imperfect wide-beam patterns. 
And for multiple stream transmissions, this process needs to be repeated several times. Therefore, its advantage of lower training overhead diminishes for mmWave cell-free MIMO with a large number of BS-UE pairs.

Recently, with the help of machine learning (ML) models, {user location \cite{heng2021machine}, \cite{va2017inverse} and sounding/probing beam measurements \cite{alkhateeb2018deep}, \cite{heng2021learning}}, are leveraged to predict the optimal beam or {several strongest beams simultaneously} for accelerating the BA procedure. 
The spatial distribution of channel power depends not only on the user location but also on the environment geometry, e.g., the position of obstacles, etc., location-based solutions are therefore mainly suitable for a line-of-sight (LOS) environment. 
In addition, sensors such as GPS or radar are required, and the location information is inaccurate in indoor environments.

Compared to the user location, sounding/probing beam measurements with quasi-omni beam pattern \cite{alkhateeb2018deep} or multi-peak beam pattern \cite{heng2021learning}, provide a multi-path signature of the surrounding propagation environment, enabling subsequent ML-based beam prediction to support the BA in both LOS and Non-LOS (NLOS) scenarios. 
Specifically, for mmWave cell-free MIMO downlink with multiple BSs coordinated by the central unit (CU), the uplink training is initially employed to reduce the pilot overhead by leveraging the uplink-and-downlink channel reciprocity.
 Subsequently, the CU aggregates the receiving signals from multiple BSs using quasi-omni beams and predicts the optimal beams for downlink transmission \cite{alkhateeb2018deep}.
 {In low SNR scenarios, the performance of quasi-omni beam-based learning degrades, similar to hierarchical beam sweeping.} 
 For point-to-point mmWave massive MIMO downlink, the pattern of multiple probing beams and the ML-based optimal-beam predictor are jointly optimized in an end-to-end manner \cite{heng2021learning}.

Most ML-aided BA approaches mentioned above rely on centralized learning.
This places high demands on the performance of the ML engine, requiring powerful hardware support, high computational power, and storage capacity. 
And the communication overhead significantly increases as the number of nodes and/or local data samples grows larger. 
In distributed learning, different nodes can collaboratively train the ML model without direct local raw-data sharing, thus reducing the communication overhead and relieving the computational and storage pressure on individual nodes \cite{chen2021distributed}.

A fully distributed uplink BF scheme based on deep reinforcement learning (DRL) was proposed in \cite{fredj2022distributed} where the CU collects learning experiences from multiple BSs to realize local BF designs. 
In \cite{hojatian2022decentralized}, fully and partially distributed unsupervised deep neural network (DNN) architectures were proposed for cell-free MIMO systems, which perform coordinated BF with zero and limited fronthaul link overhead, respectively. 
Researchers introduced a federated learning (FL) framework in \cite{elbir2020federated} to train a convolutional neural network (CNN) for hybrid BF, where the model training takes place at the BS by collecting gradients solely from multiple users.
In the distributed BA method for LIDAR-assisted mmWave systems \cite{mashhadi2021federated}, connected vehicles collaboratively train a shared DNN based on local LIDAR data under the FL framework.

% 除了分布式实现之外,模型的小样本场景适配和快时变更新问题,引出宽学习,简单说明原理和已有在应用（非无线通信与无线通信）。

In mmWave cell-free MIMO systems with fast time-varying or even non-stationary channels, the ML model for BA should be adjusted frequently, which leads to a small valid training dataset.
Additionally, the computational overhead due to frequent model updates cannot be neglected.
Broad learning (BL), based on a flat-form neural network, has shown its efficiency and effectiveness in addressing regression and classification problems \cite{chen2017broad, chen2018universal}.
In comparison to DNN, BL requires less training time while achieving comparable or even better performance for problems that have moderate learning difficulty and insufficient training data.
Furthermore, the structural feature of the BL model enables incremental updating to handle the periodic arrival of online data, showcasing its applicability to time-varying scenarios.
Recent studies \cite{long2019broad} and \cite{long2021few} have employed the BL model for mmWave hybrid BF, utilizing semi-supervised learning and few-shot learning for online implementation, despite the high cost of BF labeling and the non-stationary nature of scenarios involving the birth-death process of scattering paths.

% 文章贡献
%各用户收集下行宽波束响应与传输窄波束响应，训练本地宽学习网络，并基于预测的窄波束响应进行波束选择。进一步，通过将各用户本地网络的训练问题建模为具有一致性约束的分布式优化问题，利用相邻用户间的D2D通信，可实现训练数据的有效共享。进一步，设计协作模式下用户本地网络的增量式更新方式，可有效降低网络的训练复杂度。本发明充分利用分布式宽学习在小样本条件下挖掘多基站宽波束响应与传输窄波束响应关系的能力，可实现快时变场景多点协作毫米波大规模系统的低复杂度低开销波束选择。
This paper focuses on studying the design of {probing} beam-based BL-aided BA for mmWave cell-free MIMO downlink systems.
For scenarios without uplink-downlink channel reciprocity, e.g., frequency-division duplexing (FDD) systems, we propose a user-side BL-aided incremental collaborative BA approach, which enables effective distributed implementation of training data sharing.
During the offline phase, each user collects downlink measurements of both {probing} beams and transmission narrow beams.
Each user can utilize their own collection of multiple BS probing beam measurements to perform beam prediction.
However, the local training data of the user may be insufficient to handle fast time-varying channels.
We propose a collaborative training approach by formulating the training problem of multiple users' local BL models as a distributed optimization problem with consistency constraints.
Additionally, we propose an incremental update of the BL model in the collaborative training mode to reduce the complexity of the model update.
During the online phase, each user utilizes the updated local BL model for beam prediction based solely on the local measurements of multiple BS {probing} beams.

%For scenarios with uplink-downlink channel reciprocity, e.g., time-division duplexing (TDD) systems, we propose a BS-side BL-aided incremental collaborative BA approach, which can achieve an efficient implicit sharing of multiple BS features.
%In the offline phase, each BS collects uplink measurements of both {probing} beam and transmission narrow beams. The optimal BL model is the centralized one implemented at the CU which aggregates multiple BSs' beam measurements. To reduce the overhead of fronthaul links and the computational complexity of the CU, the training of the BL model is solved in a distributed fashion under the vertical FL framework \cite{chen2020vafl} with each BS only handling local {probing} beam measurement via splitting across features. Moreover, we design a maximum value-based sparsification method to further reduce the communication overhead by exploiting the parameter sparsity during the training process. In the online phase, each BS first conducts beam prediction based on its locally updated BL model and {probing} beam measurement. Then, these intermediate results from multiple BSs are reported and combined at the CU to make the final beam prediction.
%特征在于，针对多点协作毫米波MIMO场景的下行波束选择问题，借鉴纵向联邦学习框架，将原集中式的多基站上行宽波束响应与最佳传输窄波束的映射问题，通过垂直切割数据特征空间，转化为分布式学习问题，并设计了具体的基站协作分布式波束选择框架。同时，通过挖掘训练过程中间参数的稀疏性，降低了前传链路的通信开销。进一步，设计协作模式下基站本地网络的增量式更新方式，有效降低了网络的更新复杂度。本发明充分利用分布式宽学习在小样本条件下挖掘多基站宽波束响应与传输窄波束响应关系的能力，可实现多基站协作毫米波MIMO系统的低开销波束选择，缓解当前方案对中央处理单元单一引擎性能及前传链路带宽的较高要求

For scenarios with uplink-downlink channel reciprocity, e.g., time-division duplexing (TDD) systems, we propose a BS-side BL-aided incremental collaborative BA approach, which enables an efficient implicit sharing of multiple BS features.
During the offline phase, each BS collects uplink measurements of both {probing} beams and transmission narrow beams. The optimal BL model is implemented centrally at the CU, which aggregates the beam measurements from multiple BSs. 
To reduce the overhead of fronthaul links and the computational complexity of the CU, we solve the training of the BL model in a distributed fashion using the vertical FL framework \cite{chen2020vafl}, where each BS only handles the local {probing} beam measurement by feature splitting. 
Furthermore, we design a sparsification method based on maximum values to further reduce the communication overhead by leveraging the sparsity of parameters during the training process. 
During the online phase, each BS initially performs beam prediction based on its updated local BL model and the {probing} beam measurement. Subsequently, the CU receives and combines these intermediate results from multiple BSs to generate the final beam prediction.

In summary, our proposed user-side and BS-side incremental collaborative BL-aided BA approaches fully utilize the distributed BL's ability to explore the relationship between multiple BS {probing} beam measurements and transmission narrow beam responses.
The user-side approach is primarily targeted towards mmWave cell-free MIMO downlink systems lacking uplink-downlink channel reciprocity and having a low-to-medium valid dataset size due to fast time-varying and/or non-stationary channels.
The BS-side approach, on the other hand, is mainly designed for mmWave cell-free MIMO downlink systems that possess uplink-downlink channel reciprocity, fronthaul links with low-to-medium bandwidth, and CU with inadequate computing power.
Simulation results illustrate the advantages of our proposed approaches in comparison to traditional BA schemes and DNN-aided BA schemes for the aforementioned scenarios.

% 章节安排
The paper is organized as follows. 
Section II introduces the system model and problem formulation. 
Sections III and IV elaborate on the proposed user-side and BS-side incremental collaborative BL-aided BA approaches, respectively. 
Section V presents simulation results and related discussions. Finally, Section VI concludes the paper.

In this paper, bold upper case letters and bold lower case letters denote matrices and vectors, respectively. The conjugate transpose and transpose of $\mathbf{A}$ are denoted
by $\mathbf{A}^\text{H}$ and $\mathbf{A}^{\text{T}}$.
$\otimes$ denotes the Kronecker product. 
${\rm blkdiag}\{\cdot\}$ is the operator for block diagonal matrix.
$\Vert \cdot \Vert_\text{F}$ denotes the Frobenius norm.
$\mathcal{{CN}}\left(\text{\ensuremath{\boldsymbol{\mu}}},\boldsymbol{\Sigma}\right)$ denotes the circularly symmetric complex Gaussian distribution with mean $\boldsymbol{\mu}$ and covariance $\boldsymbol{\Sigma}$. $\angle$ takes the phase of a complex number. $\mathbf{0}_{M\times N}$ denotes a matrix with all zero elements. $\boldsymbol{\boldsymbol{1}}_{N\times1}$ and
$\mathbf{I}_{N}$ denote the $N$-dimensional all-ones vector and
identity matrix, respectively. 
 $a=\mathcal{O}\left(b\right)$
means that $a$ and $b$ have the same scaling.

\section{System Model}
\label{sec2}
%\begin{figure}[htbp]
%\centering
%\setlength{\abovecaptionskip}{0.cm}
%\includegraphics[scale=0.7]{System_model.pdf}
%\caption{System model.}\label{system model}
%\end{figure}
We consider a mmWave cell-free MIMO downlink system, where $B$ BSs each with $M$ antennas cooperatively serve $U$ single-antenna users via orthogonal frequency division multiple access (OFDMA).
The system bandwidth is $B_w$ Hz and the subcarrier number is $K$. Denote the sets of BSs, users and subcarriers as $\mathbb{B}=\{1, \ldots, B\}$, $\mathbb{U}=\{1, \ldots, U\}$ and $\mathbb{K}=\{1,\ldots,K\}$, respectively. The subset of subcarriers for user $u\in\mathbb{U}$ is $\mathbb{K}_u=\{k_{u,1},\ldots,k_{u,K_u}\}$ with $K_u=|\mathbb{K}_u|$. For practical consideration, each BS uses $U$ RF chains and hybrid BF for downlink transmission.

\subsection{Channel Model}
The downlink channel in the antenna-subcarrier domain from BS $b\in \mathbb{B}$ to user $u\in\mathbb{U}$ is
\setlength\abovedisplayskip{1pt}\setlength\belowdisplayskip{1pt}
\begin{equation}
{\mathbf h}_{b, u, k}=\sum_{l=1}^{L} \alpha_{b, u, l} e^{-j 2 \pi f_{k} \tau_{{b, u, l}}} \mathbf{a}\left(\theta_{b, u, l}, \phi_{b, u, l}\right),
\end{equation}
where $L$ denotes the number of distinguishable propagation paths. $\alpha_{b,u,l}$, $\tau_{b, u, l}$, $\theta_{b, u, l}$ and $\phi_{b, u, l}$ are the complex gain, propagation delay, azimuth and elevation angle of path $l$. $f_k$ denotes the central frequency of subcarrier $k \in \mathbb{K}_u$. For the uniform planar array (UPA) with $H$ and $W$ antennas in vertical and horizontal directions ($M=WH$), the array steering vector satisfies
$
\mathbf{a}\left(\theta_{b, u, l}, \phi_{b, u, l}\right)=\mathbf{a}_{z}\left(\phi_{b, u, l}\right) \otimes \mathbf{a}_{y}\left(\theta_{b, u, l}, \phi_{b, u, l}\right),
$
where
$
\mathbf{a}_{{z}}\left(\phi_{b, u, l}\right)=\left[1, e^{j\frac{2\pi}{\lambda} d \cos \left(\phi_{b, u, l}\right)}, \cdots, e^{j \frac{2\pi}{\lambda} d(H-1) \cos \left(\phi_{b, u, l}\right)}\right]^\text{T} \in \mathbb{C}^{H\times 1}
$
and
$
\mathbf{a}_{{y}}\left(\theta_{b, u, l}, \phi_{b, u, l}\right)=\left[1, e^{j \frac{2\pi}{\lambda} d \sin \left(\theta_{b, u, l}\right) \sin \left(\phi_{b, u, l}\right)}, \cdots \right., $ $\left. e^{j \frac{2\pi}{\lambda} d(W-1) \sin \left(\theta_{b, u, l}\right) \sin \left(\phi_{b, u, l}\right)}\right]^\text{T}\in \mathbb{C}^{W\times 1}
$
with $\lambda$ and $d$ being the downlink wavelength and antenna spacing, respectively.

\subsection{Transmission Model}
We adopt hybrid BF for downlink transmission, where analog beams are chosen from a standard Discrete Fourier Transformation (DFT) codebook $\mathbf{F}=\left[\mathbf{f}_1,...,\mathbf{f}_M\right]\in \mathbb{C}^{M\times M}$ that satisfies $\mathbf{F}^\text{H}\mathbf{F}=\mathbf{F}\mathbf{F}^\text{H}=\mathbf{I}_M$. 
Given $\mathbf{f}_{b,u}^\text{RF}=\mathbf{f}_{i_{b,u}}$ as the analog beam of BS $b$ for user $u$ with $i_{b,u}$ being the codeword index, the analog BF matrix can be represented as $\mathbf{F}_{u}^\text{RF}={\rm blkdiag}\left\{\mathbf{f}_{1,u}^\text{RF},\cdot\cdot\cdot,\mathbf{f}_{B,u}^\text{RF}\right\}$. Next, the baseband BF is designed based on the maximum-ratio-transmission (MRT) principle. Specifically, for user $u$ at subcarrier $k$, the baseband BF is $\mathbf{f}_{u,k}^\text{CU}=\frac{\left(\mathbf{h}_{u,k}^\text{H} \mathbf{F}_{u}^\text{RF}\right)^\text{H}}{\Vert \mathbf{h}_{u,k}^\text{H}\mathbf{F}_{u}^\text{RF} \Vert_\text{F}}$ with $\mathbf{h}_{u,k} =\left[\mathbf{h}_{1,u,k}^\text{T},\mathbf{h}_{2,u,k}^\text{T},\cdot\cdot\cdot,\mathbf{h}_{B,u,k}^\text{T}\right]^\text{T}$. Then the received signal of user $u$ at subcarrier $k \in \mathbb{K}_{u}$ is 
\begin{equation}
	\begin{aligned}
		y_{u, k}&=\mathbf{h}_{u, k}^{\text H} \mathbf{F}_u^{\text{RF}} \mathbf{f}_{u, k}^{\text{CU}} s_{u, k}+n_{u, k} \\
		&=\sqrt{\sum\nolimits_{b=1}^{B}\left|\mathbf{h}_{b, u, k}^\text{H} \mathbf{f}_{b,u}^\text{RF}\right|^{2}} s_{u, k}+n_{u, k}, 
	\end{aligned}
\end{equation}
where $s_{u, k} \sim \mathcal{CN}\left(0, P_{u,k}\right)$ and $n_{u, k} \sim \mathcal{CN}\left(0, \sigma^2\right)$ are the transmitted symbol and the receiver noise, respectively. Given $\left\|\mathbf{F}_{u}^\text{RF} \mathbf{f}_{u, k}^\text{CU}\right\|_\text{F}=1$ and the total transmitting power $P$, we have $P_{u,k} = \frac{P}{UK_u}$ for average power allocation over users and subcarriers. And the receiving SNR at subcarrier $k$ of user $u$ is 
$
	\rho_{u, k}=\frac{P_{u,k}\sum_{b=1}^{B}\left|\mathbf{h}_{b, u, k}^\text{H} \mathbf{f}_{b, u}^\text{RF}\right|^{2}}{\sigma^{2}}.
$
The corresponding downlink achievable rate (bit/s) provided from subcarrier $k$ for user $u$ is 
$
R_{u, k}=\frac{B_{w}}{K} \log _{2}\left(1+\rho_{u, k}\right).	
$

Note that the BS side should acquire the downlink channel state information (CSI) first for subsequent hybrid BF. For mmWave cell-free MIMO systems, this is generally based on beam training. However, training all beams with indices $i_{b,u}\in \mathbb{M}=\{1,2,...,M\}$ for all BSs $b\in\mathbb{B}$ and all users $u\in\mathbb{U}$ is very time consuming. Define the channel tracking period as $T$, i.e., the system needs to re-conduct beam training every $T$ to update the hybrid BF.
During the first $T_\text{r}$ time of each period, the BS side performs beam training and the remaining time is used for data transmission. Therefore, the effective downlink rate of user $u$ can be represented as
\setlength\abovedisplayskip{1pt}\setlength\belowdisplayskip{1pt}
\begin{equation}
	R_{u}^{\text {eff }}=\left(1-\frac{T_\text{r}}{T}\right) \frac{B_{w}}{K} \sum_{k \in \mathbb{K}_{u}} \log _{2}\left(1+\frac{P_{u,k}}{\sigma^{2}} \sum_{b=1}^{B}\left|\mathbf{h}_{b, u, k}^\text{H} \mathbf{f}_{b, u}^\text{RF}\right|^{2}\right).
\end{equation}
We can formulate the problem of beam selection as
$
\max _{i_{b,u}\in \mathbb{M}, b \in \mathbb{B}, u \in \mathbb{U}} R^{\text{eff}}=\sum_{u=1}^{U} R_{u}^{\text{eff}}.
$
Since user interference is negligible under OFDMA, the above problem is converted to 
$
	\max _{i_{b,u}\in \mathbb{M}, b \in \mathbb{B}} R_{u}^{\text{eff}},
$
for $ u \in \mathbb{U}$. However, the solution space is the cascaded codebook space of $B$ BSs with the space size of $M^B$, which is generally overwhelming for practical systems. Therefore, we use the following 
conversion, i.e.,
$
\max _{i_{b,u}\in \mathbb{M}}\left(1-\frac{T_\text{r}}{T}\right) \sum_{k \in \mathbb{K}_{u}} \log _{2}\left(1+\frac{P_{u,k}}{\sigma^{2}}\left|\mathbf{h}_{b, u, k}^\text{H} \mathbf{f}_{b, u}^\text{RF}\right|^{2}\right),	
$
for $ b \in \mathbb{B} $ and $ u \in \mathbb{U}$, to achieve a sub-optimal solution with low complexity. Here, $c_{b, u}=\sum_{k \in \mathbb{K}_{u}} \log_{2}\left(1+\frac{P_{u,k}}{\sigma^{2}}\left|\mathbf{h}_{b, u, k}^\text{H} \mathbf{f}_{b, u}^\text{RF}\right|^{2}\right)$ is the narrow-beam equivalent rate.

For the converted problem, if the complete beam-domain CSI amplitude, i.e., $\left|\mathbf{h}_{b, u, k}^\text{H} \mathbf{f}_{b, u}^\text{RF}\right|, i_{b, u} \in \mathbb{M}$, is available, we can maximize the equivalent rate of all narrow beams $c_{b, u}$'s. However, given the time for training one beam as $T_\text{b}$, this requires $T_\text{r}=BMT_\text{b}$ for orthogonal downlink training and at least $T_\text{r}=MT_\text{b}$ for orthogonal uplink training. For systems with uplink-downlink reciprocity, e.g., TDD systems, less time overhead is required for uplink training. However, downlink training should be conducted if this reciprocity does not exist, e.g., for FDD systems. For typical mmWave cell-free MIMO downlink systems with either downlink or uplink training, the time overhead $T_\text{r}$ is non-negligible due to $M\geq 1$, which may result in a low effective rate, especially for scenarios with small $T$, e.g., fast time-varying channels.

{Previous studies have demonstrated that the use of a small number of probing beams, e.g., with quasi-omni \cite{alkhateeb2018deep} or multi-peak beam pattern \cite{heng2021learning}, helps the receivers to perceive information about the environmental characteristics. 
And the probing beam responses of multiple BSs can provide an implicit representation of user's location \cite{alkhateeb2018deep}. Since the user's location directly affects the beam-domain CSI which determines the BA result, it can be inferred that there exists a mapping relationship between the probing beam responses of multiple BSs and the optimal BA.} 

\section{User-Side Incremental Collaborative BL-aided BA Design}
\begin{figure}[htp]%[!h]
%是可选项 h表示的是here在这里插入,t表示的是在页面的顶部插入
\centering
\includegraphics[scale=0.32]{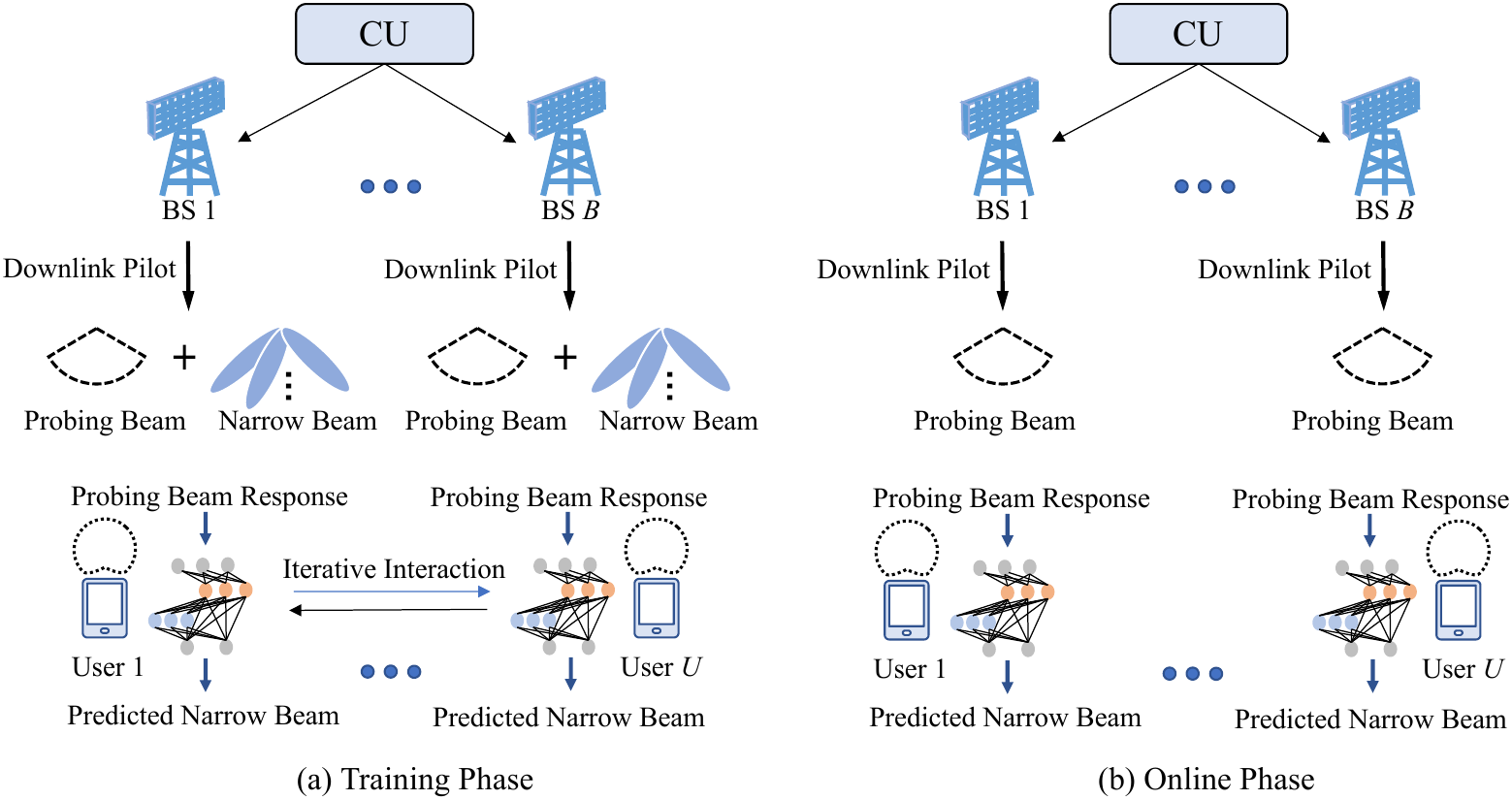}
\caption{{Execution flow of the BL-based beam prediction at the user side.}}
\label{fig:1}
\end{figure}

In this section, we present a BL-based incremental collaborative downlink BA approach for mmWave cell-free MIMO downlink without uplink-downlink channel reciprocity, where each user gathers the measurements of both {probing} beams and narrow beams during the offline phase to train their respective local BL models. These models are then utilized in the online phase to predict the optimal narrow beam based on the {probing} beam responses.
Fig. \ref{fig:1} illustrates the execution flow of the proposed approach.
To leverage the local data from multiple users concurrently and enhance the prediction accuracy, we formulate a distributed optimization problem and introduce a collaborative model training approach with low communication overhead by drawing lessons from the alternating direction method of multipliers (ADMM) method {\cite{boyd2011distributed}}. 
Finally, by capitalizing on the incremental update capability of the BL model, we put forward an incremental collaborative model training approach for fast time-varying scenarios.

\subsection{Downlink Beam Training}

Define {$\mathbf{g}_{b}^{i}, i =  -N_{\mathbf{W}}+1,...,0$} and $\mathbf{g}_{b}^{i}=\mathbf{f}_\mathit{i},i\in\mathbb{M}$ as the {$N_{\mathbf{W}}\ge 1$ probing beams} and the $i$-th narrow beam of BS $b\in\mathbb{B}$, respectively. 
The received pilot of user $u\in\mathbb{U}$ on subcarrier $k\in\mathbb{K}_{u}$ for the {$i\in\{-N_{\mathbf{W}}+1,...,M\}$}-th beam can be represented as 
$
	{y}_{b,u,k}^{i}=\mathbf{h}_{b,u,k}^\text{H}\mathbf{g}_{b}^{i}s_{b,k}^\text{pilot}+n_{u,k},
$
where $s_{b,k}^\text{pilot}=\sqrt{P_{b,k}^\text{tr}}$ is the pilot for $i$-th beam training with $P_{b,k}^\text{tr}$ being the effective training power. Therefore, user $u$ can acquire the estimation of each beam response $\mathbf{h}_{b,u,k}^\text{H}\mathbf{g}_{b}^{i}$ as 
$
	{\mathit{r}}_{b,u,k}^{i}=\mathbf{h}_{b,u,k}^\text{H}\mathbf{g}_{b}^{i}+\frac{n_{u,k}}{\sqrt{P_{b,k}^\text{tr}}}.
$
{Note that practical systems may not allow per subcarrier resolution of CSI. In addition, the beam response does not change much in consecutive subcarriers. Define $\bar{\mathbb{K}}_u=\{\bar{k}_{u,1},\ldots,\bar{k}_{u,\bar{K}_u}\}$ with $\bar{K}_u=|\bar{\mathbb{K}}_u|$ and ${\mathbb{K}}_{u}^{(\bar{k})}, \bar{k}\in \bar{\mathbb{K}}_{u}$ as the set of indices of the subcarrier groups and the set of indices of the subcarriers belonging to the group $\bar{k}$ for user $u$, respectively. Then, the beam response of subcarrier group $\bar{k}$ for user $u$ can be represented as $\hat{\mathit{r}}_{b,u,\bar{k}}^{i}=\frac{\sum_{k\in {\mathbb{K}}_{u}^{(\bar{k})}}{\mathit{r}}_{b,u,{k}}^{i}}{\left|{\mathbb{K}}_{u}^{(\bar{k})}\right|}$.}

\subsection{User-side Collaborative BL Modeling}\label{major_sec_IIIB}
In the offline phase, each user $u\in\mathbb{U}$ collects $N$ samples, i.e., {$\left\{ \mathit{\hat{r}}_{b,u,\bar{k}}^{i,\left(n\right)}\left|\right.{i= -N_{\mathbf{W}}+1,...,0},\right.$
$\left.{\bar{k}\in\bar{\mathbb{K}}_{u}},b\in\mathbb{B}\right\}$
 and $\left\{ \mathit{\hat{c}}_{b,u,i}^{\left(n\right)}=\sum\limits_{\bar{k}\in\bar{\mathbb{K}}_{u}}\left|{\mathbb{K}}_{u}^{(\bar{k})}\right|\log_{2}\left(1+\frac{\bar{P}_{u,\bar{k}}}{\sigma^{2}}\left|\hat{r}_{b,u,\bar{k}}^{i,(n)}\right|^{2}\right)\left|\right.i\in\mathbb{M}, b\in\mathbb{B}\right\}$ with $\bar{P}_{u,\bar{k}}=\frac{\sum_{k\in {\mathbb{K}}_{u}^{(\bar{k})}}{P}_{u,{k}}}{\left|{\mathbb{K}}_{u}^{(\bar{k})}\right|}$,
where $\mathit{\hat{r}}_{b,u,\bar{k}}^{i,\left(n\right)}$ and $\mathit{\hat{c}}_{b,u,i}^{\left(n\right)}$ are the $n$-th sample of $i$-th beam response for $n=1,\ldots,N$. 
 For the training of user local BL model, the $n$-th input is {$\mathbf{x}_{u}^{\left(n\right)}=\left[\mathbf{x}_{-N_{\mathbf{W}}+1,u}^{\left(n\right),{\text{T}}},\mathbf{x}_{-N_{\mathbf{W}}+2,u}^{\left(n\right),{\text{T}}},\ldots,\mathbf{x}_{0,u}^{\left(n\right),{\text{T}}}\right]^{\text{T}}\in\mathbb{R}^{2N_{\mathbf{W}}B\bar{K}_{u}\times1}$ with 
 $\mathbf{x}_{i,u}^{\left(n\right)}=\left[\mathbf{x}_{i,1,u}^{\left(n\right),\text{T}},\mathbf{x}_{i,2,u}^{\left(n\right),\text{T}},\ldots,\mathbf{x}_{i,B,u}^{\left(n\right),\text{T}}\right]^{\text{T}}\in\mathbb{R}^{2B\bar{K}_{u}\times1}$ for $i=-N_{\mathbf{W}}+1,...,0$ and
$
	\mathbf{x}_{i,b,u}^{\left(n\right)}=\left[\left|\hat{\mathit{r}}_{b,u,\bar{k}_{u,1}}^{i,\left(n\right)}\right|,\angle\hat{\mathit{r}}_{b,u,\bar{k}_{u,1}}^{i,\left(n\right)},\ldots,\left|\hat{\mathit{r}}_{b,u,\bar{k}_{u,\bar{K}_{u}}}^{i,\left(n\right)}\right|,\angle\hat{\mathit{r}}_{b,u,\bar{k}_{u,\bar{K}_{u}}}^{i,\left(n\right)}\right]^{\text{T}}\in\mathbb{R}^{2\bar{K}_{u}\times1}.
$}} 
$\hat{\mathbf{c}}_{b,u}^{\left(n\right)}\triangleq\left[\mathit{\hat{c}}_{b,u,1}^{\left(n\right)},...,\mathit{\hat{c}}_{b,u,M}^{\left(n\right)}\right]^\text{T}$. 
To reduce the learning difficulty, we adopt the one-hot coding of $\hat{\mathbf{c}}_{b,u}^{\left(n\right)}$, i.e., $\bar{\mathbf{c}}_{b,u}^{\left(n\right)}$ with
\setlength\abovedisplayskip{1pt}\setlength\belowdisplayskip{1pt}
\begin{equation}\label{Eq_15_zc1}
	\bar{{c}}_{b,u,i}^{\left(n\right)}=\begin{cases}
1, \text{ if  } \hat{{c}}_{b,u,i}^{\left(n\right)}\ge \hat{{c}}_{b,u,j}^{\left(n\right)}, \forall j\ne i,\\ 
0, \text{ otherwise,}\\
\end{cases}
\end{equation}
to construct the $n$-th label for the learning model $\mathbf{y}_{u}^{\left(n\right)}=\left[\bar{\mathbf{c}}_{1,u}^{\left(n\right),\text{T}},...,\bar{\mathbf{c}}_{B,u}^{\left(n\right),\text{T}}\right]^\text{T}\in\mathbb{R}^{BM\times1}$.
Then, the training set with $N$ samples, i.e., $\mathbf{X}_{u}\in\mathbb{R}^{N\times2{N_{\mathbf{W}}B\bar{K}_{u}}}$ and $\mathbf{Y}_{u}\in\mathbb{R}^{N\times BM}$, is
\setlength\abovedisplayskip{1pt}\setlength\belowdisplayskip{1pt}
\begin{equation}
\mathbf{X}_{u}=\left[\mathbf{x}_{u}^{\left(1\right)},\ldots,\mathbf{x}_{u}^{\left(N\right)}\right]^{\text{T}}, \text{    } \text{    } \mathbf{Y}_{u}=\left[\mathbf{y}_{u}^{\left(1\right)},\ldots,\mathbf{y}_{u}^{\left(N\right)}\right]^{\text{T}}.
\end{equation}

Recall that the BL model is established as a flat
network, where the original inputs are transferred and positioned as
``mapped features" in the feature nodes, while the structure is expanded
broadly in the ``enhancement nodes" \cite{chen2017broad}.
The training process of a BL network consists of two primary stages: 1) random generation of weights for mapped features and enhancement features, and 2) computation of weights between the hidden layer and output layer.

\begin{figure}[!h]
%是可选项 h表示的是here在这里插入,t表示的是在页面的顶部插入
\centering
\includegraphics[scale=0.35]{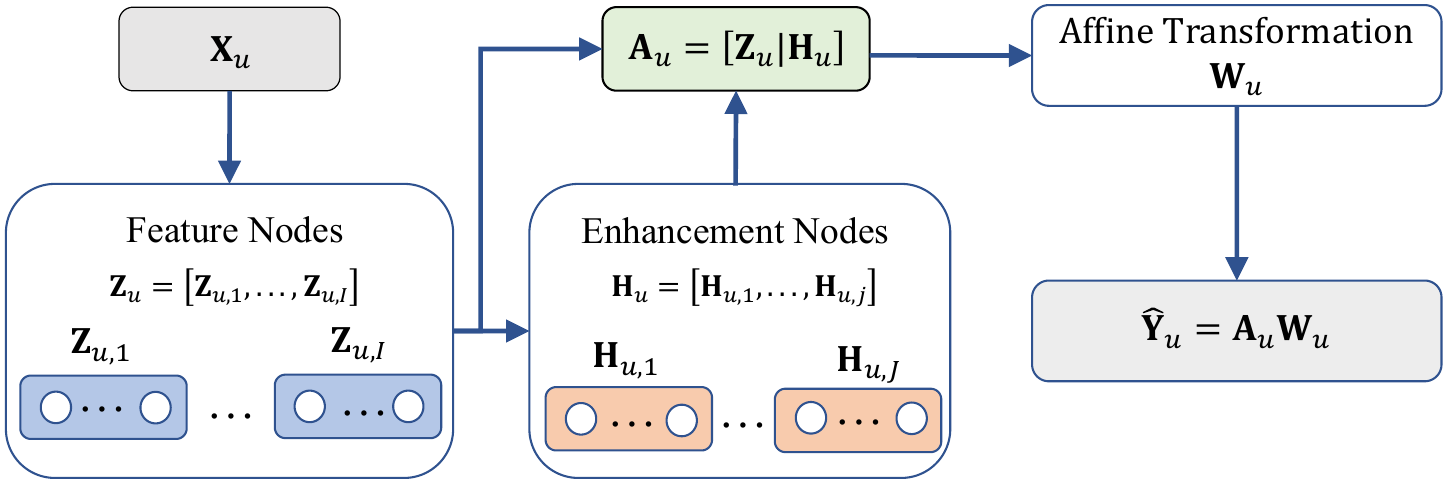}
\caption{The BL model for beam prediction.}
\label{fig:2}

\end{figure}

If each user trains the BL model for beam prediction locally, 
as depicted in Fig. \ref{fig:2}, 
it first maps the input $\mathbf{X}_{u}$ to $I$ groups of feature nodes $\mathbf{Z}_{u,i}\in\mathbb{R}^{N\times F}$, $i=1,\ldots,I$, i.e.,
\begin{equation}
\mathbf{Z}_{u,i}=\mathit{\phi}\left(\mathbf{X}_{u}\mathbf{W}_{u,e_{i}}+\mathbf{1}_{N\times1}\boldsymbol{\beta}_{u,e_{i}}\right),\label{eq_18_zc1}
\end{equation}
where $\mathbf{W}_{u,e_{i}}\in\mathbb{R}^{2{{N_{\mathbf{W}}}B\bar{K}_{u}}\times F}$ and $\boldsymbol{\beta}_{u,e_{i}}\in\mathbb{R}^{1\times F}$ are the connection weight matrix and bias vector of the feature generation layer, respectively.  These feature nodes are further mapped into $J$ groups of enhancement nodes $\mathbf{H}_{u,j}\in\mathbb{R}^{N\times E}$, $j=1,\ldots,J$, i.e., 
\setlength\abovedisplayskip{1pt}\setlength\belowdisplayskip{1pt}
\begin{equation}
\mathbf{H}_{u,j}=\mathit{\xi}\left(\mathbf{Z}_{u}\mathbf{W}_{u,h_{j}}+\mathbf{1}_{N\times1}\boldsymbol{\beta}_{u,h_{j}}\right),\label{eq_19_zc1}
\end{equation}
where $\mathbf{Z}_{u}=\left[\mathbf{Z}_{u,1},\mathbf{Z}_{u,2},\ldots,\mathbf{Z}_{u,I}\right]$ is the cascade matrix of $I$ groups of feature nodes.
$\mathbf{W}_{u,h_{j}}\in\mathbb{R}^{IF\times E}$ and $\boldsymbol{\beta}_{u,h_{j}}\in\mathbb{R}^{1\times E}$ are the connection weight matrix and bias vector of the feature enhancement layer.
The activation function $\mathit{\phi}\left(\cdot\right)$ can be either linear or nonlinear while $\xi\left(\cdot\right)$ is generally nonlinear.
In the BL framework, we randomly create weight matrices, i.e., $\mathbf{W}_{u,e_{i}}$'s, $\mathbf{W}_{u,h_{j}}$'s, along with bias vectors, i.e., $\boldsymbol{\beta}_{u,e_{i}}$'s and $\boldsymbol{\beta}_{u,h_{j}}$'s. These matrices and vectors are not trainable. 
Subsequently, we process the feature nodes and enhancement nodes $\mathbf{A}_{u}=\left[\mathbf{Z}_{u}\mid\mathbf{H}_{u}\right]\in\mathbb{R}^{N\times\left(IF+JE\right)}$ with $\mathbf{H}_{u}=\left[\mathbf{H}_{u,1},\mathbf{H}_{u,2},\ldots,\mathbf{H}_{u,J}\right]$ using an affine transformation $\mathbf{W}_{u}\in\mathbb{R}^{\left(IF+JE\right)\times BM}$ to produce the output
$\hat{\mathbf{Y}}_{u}=\mathbf{A}_{u}\mathbf{W}_{u}\in\mathbb{R}^{ N \times BM}$.
The optimization of $\mathbf{W}_{u}$ can be formulated as
\setlength\abovedisplayskip{1pt}\setlength\belowdisplayskip{1pt}
\begin{equation}\label{eqn_opt_320}
\min\limits_{\mathbf{W}_{u}}\text{ }\frac{1}{2}\left\Vert \mathbf{Y}_{u}-\mathbf{A}_{u}\mathbf{W}_{u}\right\Vert _\text{F}^{2}+\mathit{\lambda}\left\Vert \mathbf{W}_{u}\right\Vert _\text{F}^{2},
\end{equation}
where we utilize the minimum mean square error (MMSE) criterion and the $L2$ regularization to enhance the network generalization performance. The solution of the problem \eqref{eqn_opt_320} is
\setlength\abovedisplayskip{1pt}\setlength\belowdisplayskip{1pt}
\begin{equation}\label{FDBL_W_cal}
\mathbf{W}_{u}=\lim_{\mathit{\lambda}\rightarrow0}\left(\mathit{\lambda}\mathbf{I}_{IF+JE}+\mathbf{A}_{u}^{\text{T}}\mathbf{A}_{u}\right)^{-1}\mathbf{A}_{u}^{\text{T}}\mathbf{Y}_{u}.
\end{equation}

When the propagation environments of different users are statistically similar, e.g., when users have similar movement areas, one can gather data from multiple users to train a model that has better generalization performance, particularly in scenarios with limited user local training data.
Without loss of generality, we formulate the following problem 
\setlength\abovedisplayskip{1pt}\setlength\belowdisplayskip{1pt}
\begin{equation}\label{eqn_opt_322}
\min\limits_{\mathbf{W}}\text{ }\frac{1}{2}\left\Vert \mathbf{Y}-\mathbf{A}\mathbf{W}\right\Vert _\text{F}^{2}+\lambda\left\Vert \mathbf{W}\right\Vert _\text{F}^{2},
\end{equation}
where $\mathbf{Y}=\left[\mathbf{Y}_{1}^\text{T},\cdots,\mathbf{Y}_{U}^\text{T}\right]^{\text{T}}$ and $\mathbf{A}=\left[\mathbf{A}_{1}^\text{T},\cdots,\mathbf{A}_{U}^\text{T}\right]^{\text{T}}$
to train a shared model for all users, i.e., $\mathbf{W}_u\rightarrow\mathbf{W},\forall u\in \mathbb{U}$.
The solution to problem \eqref{eqn_opt_322} is
\begin{equation}\label{eqn_opt_cen_solu}
	\mathbf{W}=\lim_{\mathit{\lambda}\rightarrow0}\left(\mathit{\lambda}\mathbf{I}_{IF+JE}+\mathbf{A}^{\text{T}}\mathbf{A}\right)^{-1}\mathbf{A}^{\text{T}}\mathbf{Y}.
\end{equation}
However, aggregating training data across users to perform centralized processing of \eqref{eqn_opt_cen_solu} on a single node, e.g., one user or the BS side, leads to significant communication overhead.
An alternative approach is to formulate an equivalent and distributed executable problem, i.e., 
\setlength\abovedisplayskip{1pt}\setlength\belowdisplayskip{1pt}
\begin{equation}\label{eqn_opt_324}
\begin{array}{c}
\underset{\mathbf{W}_u,u\in\mathbb{U}}{\min}\text{ }\frac{1}{2}\sum_{u=1}^{U}\left\Vert \mathbf{Y}_{u}-\mathbf{A}_{u}\mathbf{W}_{u}\right\Vert _\text{F}^{2}+\lambda\left\Vert \mathbf{W}_{0}\right\Vert _\text{F}^{2}\\
\text{s.t.}\text{ }\mathbf{W}_{u}-\mathbf{W}_{0}=\mathbf{0}_{\left(IF+JE\right)\times BM},u\in\mathbb{U}
\end{array},
\end{equation}
where an auxiliary matrix $\mathbf{W}_{0}\in\mathbb{R}^{\left(IF+JE\right)\times BM}$ is introduced for model consistency.
This is a global variable consensus optimization. 
%Consensus problems have a long history, especially in conjunction with ADMM which can solve optimization problems by breaking them into smaller pieces, each of which is then easier to handle \cite{boyd2011distributed}.
Consensus problems have been studied extensively, particularly in combination with ADMM, which breaks down optimization problems into smaller pieces, making them easier to handle \cite{boyd2011distributed}.

Via drawing lessons from the ADMM algorithm, an iterative and interactive solving process of problem \eqref{eqn_opt_324} can be conducted as in the following corollary. 
\begin{corollary}\label{Major1_cor1}
	For the $t$-th iteration, {via introducing a dual variable $\mathbf{O}_{u} \in\mathbb{R}^{\left(IF+JE\right)\times BM}$}, user $u\in \mathbb{U}$ should conduct 
	\setlength\abovedisplayskip{1pt}\setlength\belowdisplayskip{1pt}
	\begin{align}
		\mathbf{W}_{u}\left(t\right)&=\left(\mathbf{A}_{u}^{\text{T}}\mathbf{A}_{u}+\rho\mathbf{I}_{IF+JE}\right)^{-1}\left[\mathbf{A}_{u}^{\text{T}}\mathbf{Y}_{u}-\rho\left(\mathbf{O}_{u}\left(t-1\right) \right.\right. \nonumber \\
		&\hspace{4cm}\left.\left.-\mathbf{W}_{0}\left(t-1\right) \right)\right], \label{eqn_opt_335} \\
		\mathbf{W}_{0}\left(t\right)&=\frac{U\rho}{2\lambda+U\rho}\left(\overline{\mathbf{W}}\left(t\right)+\overline{\mathbf{O}}\left(t-1\right)\right), \label{eqn_opt_336}\\
		\mathbf{O}_{u}\left(t\right)&=\mathbf{O}_{u}\left(t-1\right)+\mathbf{W}_{u}\left(t\right)-\mathbf{W}_{0}\left(t\right),\label{eqn_opt_337}
	\end{align}
where $\overline{\mathbf{W}}(t)=\frac{1}{U}\sum_{u=1}^{U}\mathbf{W}_{u}(t)\in\mathbb{R}^{\left(IF+JE\right)\times BM}$, $\overline{\mathbf{O}}(t)=\frac{1}{U}\sum_{u=1}^{U}\mathbf{O}_{u}(t)$, and $\rho>0$ is the penalty coefficient that controls the consistency constraint. 
\end{corollary}
\begin{IEEEproof}
From \cite[Eq. (7.6)-(7.8)]{boyd2011distributed}, we have
\setlength\abovedisplayskip{1pt}\setlength\belowdisplayskip{1pt}
\begin{align}
		\mathbf{W}_{u}(t)&=\arg\text{}\min \left(\frac{1}{2}\left\|\mathbf{Y}_{u}-\mathbf{A}_{u} \mathbf{W}_{u}(t)\right\|_\text{F}^{2} \right. \nonumber
		\\ &\hspace{0.2cm}\left.+\frac{\rho}{2}\left\|\mathbf{W}_{u}(t)-\mathbf{W}_{0}(t-1)+\mathbf{O}_{u}(t-1)\right\|_\text{F}^{2}\right), \label{Eq_27zc}\\
		\mathbf{W}_{0}(t)&=\arg \text{}\min\left(\lambda\left\|\mathbf{W}_{0}(t)\right\|_\text{F}^{2}\right. \nonumber
		\\& \hspace{0.2cm}\left. +\frac{U \rho}{2}\left\|\mathbf{W}_{0}(t)-\overline{\mathbf{W}}({t})-\overline{\mathbf{O}}(t-1)\right\|_\text{F}^{2}\right), \label{Eq_28_zc}
\end{align}
and Eq. \eqref{eqn_opt_337}. Then, via taking partial derivatives of the objective functions in Eq. \eqref{Eq_27zc} and Eq. \eqref{Eq_28_zc}  with respect to $\mathbf{W}_{u}(t)$ and $\mathbf{W}_{0}(t)$ respectively, Eq. \eqref{eqn_opt_335} and Eq. \eqref{eqn_opt_336} can be obtained.
\end{IEEEproof}

In the proposed collaborative BL-aided BA design, for each user to acquire its local BL model, the cost is analyzed as follows. 
First, according to Eq. \eqref{eq_18_zc1}-\eqref{eq_19_zc1}, calculating feature nodes and enhancement nodes need $\mathcal{O}\left(2NN_{W}B\bar{K_{u}}IF\right)$ and $\mathcal{O}\left(NIFJE\right)$ multiplications, respectively. Denote $t_\text{max}$ as the maximum iteration number. Then, Eq. \eqref{eqn_opt_335} needs $\mathcal{O}\left(\left(IF+JE\right)^{2}N\right)+\mathcal{O}\left((IF+JE)^{3}\right)+\mathcal{O}\left((IF+JE)NBM\right)+t_\text{max}\mathcal{O}\left((IF+JE)^{2}BM\right)$ multiplications.
The overall computational complexity per user is 
$
\mathcal{O}\left(2NN_{W}B\bar{K_{u}}IF\right)+\mathcal{O}\left(NIFJE\right)+\mathcal{O}\left((IF+JE)^{2}N\right)
$
$+\mathcal{O}\left((IF+JE)^{3}\right)+\mathcal{O}\left((IF+JE)NBM\right)+t_\text{max}\mathcal{O}\left((IF+JE)^{2}BM\right)
$.
%\begin{align}
%\mathcal{O}\left(2NN_{W}B\bar{K_{u}}IF\right)+&\mathcal{O}\left(NIFJE\right)+\mathcal{O}\left((IF+JE)^{2}N\right)\nonumber \\
%+&\mathcal{O}\left((IF+JE)^{3}\right)+\mathcal{O}\left((IF+JE)NBM\right)+t_\text{max}\mathcal{O}\left((IF+JE)^{2}BM\right). \nonumber
%\end{align}
The communication overhead results from the calculation of $\overline{\mathbf{W}}(t)$ and $\overline{\mathbf{O}}(t-1)$.
Two types of communication can support these calculations. First, users can exchange their local ${\mathbf{W}}_u(t)$ and ${\mathbf{O}}_u(t-1)$ via the D2D protocol \cite{chen2017feedback}. The communication overhead per user (measured by the number of real numbers to be transferred) is $2t_\text{max}(IF+JE)BM(U-1)$.
Second, the BS side collects ${\mathbf{W}}_u(t)$ and ${\mathbf{O}}_u(t-1)$ from all users $u\in\mathbb{U}$ and then broadcasts $\overline{\mathbf{W}}(t)$ and $\overline{\mathbf{O}}(t-1)$ to them. 
The communication overhead per user is $\frac{2t_\text{max}(IF+JE)BM(U+1)}{U}$. 

In contrast, the overhead and complexity per user of local model training without user cooperation, i.e., Eq. \eqref{eq_18_zc1}, \eqref{eq_19_zc1}, \eqref{FDBL_W_cal}, are zero and  
$\mathcal{O}\left(2NN_{W}B\bar{K_{u}}IF\right)+\mathcal{O}\left(NIFJE\right)+\mathcal{O}\left((IF+JE)^{2}N\right)+
\mathcal{O}\left((IF+JE)^{3}\right)+\mathcal{O}\left((IF+JE)NBM\right)$.
The per-user overhead of centralized training based on data aggregation, i.e., Eq. \eqref{eq_18_zc1}, \eqref{eq_19_zc1}, \eqref{eqn_opt_cen_solu}, is about $N(IF+JE+BM)+\frac{(IF+JE)BM}{U}$, and its per-user complexity is $\mathcal{O}\left(2NN_{W}B\bar{K_{u}}IF\right)+\mathcal{O}\left(NIFJE\right)+\mathcal{O}\left((IF+JE)^{2}N\right)
+\frac{1}{U}\mathcal{O}\left((IF+JE)^{3}\right)+\mathcal{O}\left((IF+JE)NBM\right)$. By comparison, we know that if the iteration number $t_\text{max}$ is relatively small, when $IF+JE \gg BM$ and $N\gg BM$, i.e., the scenario of interest in our simulation, the collaborative training significantly saves the communication overhead compared to the centralized training. But the cost is that for a not-that-large number of cooperation users, the collaborative training requires additional computational complexity of at most $t_\text{max}\mathcal{O}\left((IF+JE)^{2}BM\right)$. In addition, more local data storage is needed.

In the online execution phase, $B$ BSs first send downlink pilots for {probing} beam training with time cost $B{N_{\mathbf{W}}}T_\text{b}$. Then, each user $\forall u\in\mathbb{U}$ uses the {probing} beam responses of multiple BSs, i.e., $\mathbf{x}_{u}\in\mathbb{R}^{2{N_{\mathbf{W}}B\bar{K}_{u}}\times1}$, to obtain the joint output of feature and enhancement nodes, i.e., $\mathbf{a}_{u}=\left[\mathbf{z}_{u}\mid\mathbf{h}_{u}\right]\in\mathbb{R}^{1\times\left(IF+JE\right)}$, where $\mathbf{z}_{u}=\left[\mathbf{z}_{u,1},\mathbf{z}_{u,2},\ldots,\mathbf{z}_{u,I}\right]\in\mathbb{R}^{1\times IF}$
with $\mathbf{z}_{u,i}=\mathit{\phi}\left(\mathbf{x}_{u}^\text{T}\mathbf{W}_{u,e_{i}}+\boldsymbol{\beta}_{u,e_{i}}\right)\in\mathbb{R}^{1\times F}$ and $\mathbf{h}_{u}=\left[\mathbf{h}_{u,1},\mathbf{h}_{u,2},\ldots,\mathbf{h}_{u,J}\right]\in\mathbb{R}^{1\times JE}$ with 
$\mathbf{h}_{u,j}=\mathit{\xi}\left(\mathbf{z}_{u}\mathbf{W}_{u,h_{j}}+\boldsymbol{\beta}_{u,h_{j}}\right)\in\mathbb{R}^{1\times E}$. These nodes are further processed via the trained affine transformation $\mathbf{W}_{u}$ to output the predicted narrow beam index $I^{\star}_{b,u}=\arg \max_{i\in{\mathbb{M}}}\{\hat{{y}}_{u,(b-1)M+i}\}$ for each BS $b\in \mathbb{B}$ where $\hat{\mathbf{y}}_{u}^\text{T}=\mathbf{a}_{u}\mathbf{W}_{u}\in\mathbb{R}^{1\times BM}$. {Finally, each BS $b\in \mathbb{B}$ trains the predicted beam $I^{\star}_{b,u}$ for each user $u\in\mathbb{U}$, based on which the baseband BF at the CU is conducted.}
{Note that in the online execution phase of the proposed scheme, there is no data interaction between users. The per user computational complexity is $\mathcal{O}\left(2{N_{\mathbf{W}}B{\bar{K}_{u}}IF}\right) + \mathcal{O}\left(IFJE\right)+\mathcal{O}\left((IF+JE)BM\right)$}.

Beam training during the online execution phase incurs two types of overhead.
One part of the overhead arises from training probing beams, i.e., $B{N_{\mathbf{W}}}T_\text{b}$, while the other part arises from training predicted narrow beams, i.e., $BT_\text{b}$. 
Note that the training of narrow beams for different users from the same BS can be carried out simultaneously in the OFDMA mode by assigning a dedicated RF chain for each user. 
Therefore, the total time required for beam training in each channel tracking period is $T^\text{BL}_\text{r}=B{N_{\mathbf{W}}}T_\text{b}+BT_\text{b}$.
And the effective rate of user $u$ {in the online execution phase} can be calculated as
\setlength\abovedisplayskip{1pt}\setlength\belowdisplayskip{1pt}
\begin{eqnarray}
	&\hspace{-5cm}R_{u}^{\text {eff }}=\left(1-\frac{{T^\text{BL}_\text{r}}}{T}\right) \frac{B_{w}}{K} \times \nonumber 
	\\ & \hspace{1cm}  \sum_{k \in \mathbb{K}_{u}} \log _{2}\left(1+\frac{P_{u,k}}{\sigma^{2}} \sum_{b=1}^{B}\left|\mathbf{h}_{b, u, k}^\text{H} \mathbf{f}_{I^{\star}_{b,u}}\right|^{2}\right).
\end{eqnarray}

\subsection{Incremental Model Updating}

When there are changes in the wireless environment, e.g., a large range of user movement or the movement of scatterers (e.g., cars), updating the BL model for beam prediction becomes necessary. 
In the following, we present the incremental update mechanism for the collaborative training scheme proposed above, aiming to enhance the efficiency of model updates.

Suppose that user $u\in\mathbb{U}$ collects $\acute{N}$ new samples
$\left\{ \mathbf{X}_{u}^\text{a}\in\mathbb{R}^{\acute{N}\times 2{N_{\mathbf{W}}B\bar{K}_{u}}},\mathbf{Y}_{u}^\text{a}\in\mathbb{R}^{\acute{N}\times BM}\right\} $ for the updating of its BL model. By referring to Eq. \eqref{eq_18_zc1} and Eq. \eqref{eq_19_zc1}, we can form features and enhancement nodes corresponding to these new data as $\mathbf{A}_{u}^\text{a}\in\mathbb{R}^{\acute{N}\times\left(IF+JE\right)}$. 
One approach to calculate the updated affine transformation $\mathbf{W}_{u}^\text{new}\in\mathbb{R}^{\left(IF+JE\right)\times BM}$ is to perform a certain number of iterations in Corollary \ref{Major1_cor1} using both historical and newly collected samples $\mathbf{A}_{u}^\text{S}=\left[\mathbf{A}_{u}^{\text{T}},\mathbf{A}_{u}^{\text{a},\text{T}}\right]^{\text{T}}\in\mathbb{R}^{\left(N+\acute{N}\right)\times\left(IF+JE\right)}$. 
The complexity mainly arises from the inversion operation in Eq. \eqref{eqn_opt_335}, i.e., $\mathbf{C}_{u}^\text{S}\triangleq\left(\mathbf{A}_{u}^{\text{S},\text{T}}\mathbf{A}_{u}^\text{S}+\rho\mathbf{I}_{IF+JE}\right)^{-1}\in\mathbb{R}^{\left(IF+JE\right)\times\left(IF+JE\right)}$. To handle this, we utilize the existing result $\mathbf{C}_{u}\triangleq\left(\mathbf{A}_{u}^{\text{T}}\mathbf{A}_{u}+\rho\mathbf{I}_{IF+JE}\right)^{-1}$ and propose a more efficient incremental calculation of $\mathbf{C}_{u}^\text{S}$ in the following corollary. 
% 给出计算量之后，说一下针对实时性要求很高的场景，如快时变非平稳场景，这可能会带来挑战。
\begin{corollary}\label{corollary2_zc}
The inversion result $\mathbf{C}_{u}^\text{S}$ can be calculated according to
\setlength\abovedisplayskip{1pt}\setlength\belowdisplayskip{1pt} 
\begin{equation}\label{eq_30_zc1}
	\mathbf{C}_{u}^\text{S}
	=\mathbf{C}_{u}-\mathbf{C}_{u}\mathbf{A}_{u}^{\text{a},\text{T}}\left(\mathbf{I}_{\acute{N}}+\mathbf{A}_{u}^\text{a}\mathbf{C}_{u}\mathbf{A}_{u}^{\text{a},\text{T}}\right)^{-1}\mathbf{A}_{u}^\text{a}\mathbf{C}_{u}.    
\end{equation}
\end{corollary}

\begin{IEEEproof}
	From the Woodbury Identity Equation for the matrix inversion
\cite{tylavsky1986generalization}, we have
\setlength\abovedisplayskip{1pt}\setlength\belowdisplayskip{1pt} 
\begin{equation}\label{eq_31_zc1}
\left(\mathbf{D-UBV}\right)^{-1}=\mathbf{D}^{-1}+\mathbf{D}^{-1}\mathbf{U}\left(\mathbf{B}^{-1}-\mathbf{V}\mathbf{D}^{-1}\mathbf{U}\right)^{-1}\mathbf{V}\mathbf{D}^{-1},
\end{equation}
where $\mathbf{D}$ and $\mathbf{B}$ should be non-singular. Via defining $\mathbf{D}\triangleq\mathbf{A}_{u}^{\text{T}}\mathbf{A}_{u}+\rho\mathbf{I}_{IF+JE}=\mathbf{C}_{u}^{-1}$, $\mathbf{U}\triangleq \mathbf{A}_{u}^{\text{a},\text{T}}$, $\mathbf{B}\triangleq -\mathbf{I}_{\acute{N}}$ and $\mathbf{V} \triangleq \mathbf{A}_{u}^\text{a}$, Eq. \eqref{eq_30_zc1} can be obtained form Eq. \eqref{eq_31_zc1}.	
\end{IEEEproof}

The overall complexity of Eq. \eqref{eq_30_zc1} is 
$\mathcal{O}\left(\acute{N}^3\right)+\mathcal{O}\left(\acute{N}^2\left(IF+JE\right)\right)+\mathcal{O}\left(\acute{N}(IF+JE)^2\right)$.
Compared to the direct calculation of $\left(\mathbf{A}_{u}^{\text{S},\text{T}}\mathbf{A}_{u}^\text{S}+\rho\mathbf{I}_{IF+JE}\right)^{-1}$ with complexity $\mathcal{O}\left(\left(IF+JE\right)^3\right)+\mathcal{O}\left(\left(IF+JE\right)^2\left(N+\acute{N}\right)\right)$, the incremental update saves non-negligible computational overhead, when $\acute{N}\ll IF+JE$. {For scenarios with fast time-varying channels, it is difficult to collect many valid data samples within the time granularity of model updates. Therefore, the above proposed incremental model updating works in these scenarios.}

{Given more training data, we should appropriately increase the number of parameters, e.g., the number of enhancement nodes, in the BL model to achieve a better compromise between the fitting ability and generalization ability.}
{Suppose that we add $\acute{J}$ groups of enhancement nodes. Each group has $\acute{E}$ enhancement nodes.}
With randomly created connection matrix and bias vector, i.e., $\mathbf{W}_{u,h_{j}}\in\mathbb{R}^{IF\times \acute{E}}$ and $\boldsymbol{\beta}_{u,h_{j}}\in\mathbb{R}^{1\times \acute{E}}$ for $j=J+1,...,J+\acute{J}$,
 $\acute{J}\acute{E}$ new enhancement nodes form $\mathbf{H}_{u}^\text{a}\in \mathbb{R}^{\left(N+\acute{N}\right)\times\acute{J}\acute{E}}$ according to Eq. \eqref{eq_18_zc1} and Eq. \eqref{eq_19_zc1}.
The set of all feature and enhancement nodes is then $\mathbf{A}_{u}^\text{SE}=\left[\mathbf{A}_{u}^\text{S},\mathbf{H}_{u}^\text{a}\right]\in \mathbb{R}^{\left(N+\acute{N}\right)\times\left(IF+JE+\acute{J}\acute{E}\right)}$.
{To accelerate the calculation of new weight matrix $\mathbf{W}_{u}^\text{new}\in\mathbb{R}^{\left(IF+JE+\acute{J}\acute{E}\right)\times BM}$, based on $\mathbf{C}_{u}^\text{S}$ calculated from Eq. \eqref{eq_30_zc1}, one can solve $\mathbf{C}_{u}^\text{SE}\triangleq\left(\mathbf{A}_{u}^\text{SE,\text{T}}\mathbf{A}_{u}^\text{SE}+\rho\mathbf{I}_{IF+JE+\acute{J}\acute{E}}\right)^{-1}\in\mathbb{R}^{\left(IF+JE+\acute{J}\acute{E}\right)\times\left(IF+JE+\acute{J}\acute{E}\right)}$ in the following manner.} 
\begin{corollary}\label{corollary3_zc}
The inversion result $\mathbf{C}_{u}^\text{SE}$ can be calculated according to
\setlength\abovedisplayskip{1pt}\setlength\belowdisplayskip{1pt}
\begin{equation}\label{eq_32_zc1}
	\mathbf{C}_{u}^\text{SE}
=\left[\begin{array}{ll}
\mathbf{C}_{u}^\text{S}+\mathbf{C}_{u}^\text{S}{\mathbf{A}_{u}^\text{S,\text{T}}\mathbf{H}_{u}^\text{a}}\mathbf{N}\mathbf{H}_{u}^\text{a,T} \mathbf{A}_{u}^\text{S}\mathbf{C}_{u}^\text{S} & -\mathbf{C}_{u}^\text{S}{\mathbf{A}_{u}^\text{S,\text{T}}\mathbf{H}_{u}^\text{a}}\mathbf{N} \\
-\mathbf{N} \mathbf{H}_{u}^\text{a,T} \mathbf{A}_{u}^\text{S}\mathbf{C}_{u}^\text{S} & \mathbf{N} 
\end{array}\right],
\end{equation}	
where $\mathbf{N} \triangleq \left(\rho \mathbf{I}_{\acute{J}\acute{E}}+\mathbf{H}_{u}^\text{a,T} \mathbf{H}_{u}^\text{a}-\mathbf{H}_{u}^\text{a,T} \mathbf{A}_{u}^\text{S}\mathbf{C}_{u}^\text{S}\mathbf{A}_{u}^\text{S,\text{T}} \mathbf{H}_{u}^\text{a}\right)^{-1} \in\mathbb{R}^{\acute{J}\acute{E}\times\acute{J}\acute{E}}$.
\end{corollary}

\begin{IEEEproof}
First, we have
\setlength\abovedisplayskip{1pt}\setlength\belowdisplayskip{1pt}
\begin{equation}
	\begin{aligned}
			\mathbf{A}_{u}^\text{SE,\text{T}}\mathbf{A}_{u}^\text{SE} & +\rho\mathbf{I}_{IF+JE+\acute{J}\acute{E}} = \\
		& \left[\begin{array}{cc}
			\underbrace{\rho \mathbf{I}_{I F+J E}+\mathbf{A}_{u}^\text{S,\text{T}} \mathbf{A}_{u}^\text{S}}_{\mathbf{G}} & \underbrace{\mathbf{A}_{u}^\text{S,\text{T}} \mathbf{H}_{u}^\text{a}}_{\mathbf{M}} \\
			\underbrace{\mathbf{H}_{u}^\text{a,T} \mathbf{A}_{u}^\text{S}}_{\mathbf{J}} & \underbrace{\rho \mathbf{I}_{\acute{J}\acute{E}}+\mathbf{H}_{u}^\text{a,T} \mathbf{H}_{u}^\text{a}}_{\mathbf{L}}
		\end{array}\right].
	\end{aligned}
\end{equation}
Via some simple transformations of the inverse expression in \cite[Section 9.1.3]{petersen3274matrix}, we have Eq. \eqref{eq_34_zc1} at the top of the next page. Then, Eq. \eqref{eq_32_zc1} can be derived from Eq. \eqref{eq_34_zc1}.
\setlength\abovedisplayskip{1pt}\setlength\belowdisplayskip{1pt}
\begin{figure*}
	\begin{equation}\label{eq_34_zc1}
		\left[\begin{array}{ll}
			\mathbf{G} & \mathbf{M} \\
			\mathbf{J} & \mathbf{L}
		\end{array}\right]^{-1}=\left[\begin{array}{ll}
			\mathbf{G}^{-1}+\mathbf{G}^{-1} \mathbf{M}\left(\mathbf{L}-\mathbf{J} \mathbf{G}^{-1} \mathbf{M}\right)^{-1} \mathbf{J}\mathbf{G}^{-1} & -\mathbf{G}^{-1} \mathbf{M}\left(\mathbf{L}-\mathbf{J} \mathbf{G}^{-1} \mathbf{M}\right)^{-1} \\
			-\left(\mathbf{L}-\mathbf{J} \mathbf{G}^{-1} \mathbf{M}\right)^{-1} \mathbf{J}\mathbf{G}^{-1} & \left(\mathbf{L}-\mathbf{J} \mathbf{G}^{-1} \mathbf{M}\right)^{-1}
		\end{array}\right].
	\end{equation}
\end{figure*}
\end{IEEEproof}
 
The overall complexity of Eq. \eqref{eq_32_zc1} is
%$
%	\mathcal{O}\left(\left(\acute{J}\acute{E}\right)^3\right)&+\mathcal{O}\left(\left(\acute{J}\acute{E}\right)^2\left(IF+JE\right)\right)+\mathcal{O}\left(\left(\acute{J}\acute{E}\right)\left(IF+JE\right)^2\right)\\
%	&+\mathcal{O}\left(\left(\acute{J}\acute{E}\right)^2\left(N+\acute{N}\right)\right)+\mathcal{O}\left(\left(\acute{J}\acute{E}\right)\left(N+\acute{N}\right)\left(IF+JE\right)\right) \nonumber.
%$
$\mathcal{O}\left(\left(\acute{J}\acute{E}\right)^3\right)
+\mathcal{O}\left(\left(\acute{J}\acute{E}\right)^2\left(IF+JE\right)\right)+\mathcal{O}\left(\left(\acute{J}\acute{E}\right)\left(IF+JE\right)^2\right)$
$+\mathcal{O}\left(\left(\acute{J}\acute{E}\right)^2\left(N+\acute{N}\right)\right)+\mathcal{O}\left(\left(\acute{J}\acute{E}\right)\left(N+\acute{N}\right)\left(IF+JE\right)\right) \nonumber.$
The complexity of directly calculating $\left(\mathbf{A}_{u}^\text{SE,\text{T}}\mathbf{A}_{u}^\text{SE}+\rho\mathbf{I}_{IF+JE+\acute{J}\acute{E}}\right)^{-1}$ is $\mathcal{O}\left(\left(IF+JE+\acute{J}\acute{E}\right)^3\right) + \mathcal{O}\left(\left(IF+JE+\acute{J}\acute{E}\right)^2\left(N+\acute{N}\right)\right)$.
%$\left(\mathbf{A}_{u}^\text{SE,\text{T}}\mathbf{A}_{u}^\text{SE}+\rho\mathbf{I}_{IF+JE+\acute{J}\acute{E}}\right)^{-1}$ is $\mathcal{O}\left(\left(IF+JE+\acute{J}\acute{E}\right)^3\right)+\mathcal{O}\left(\left(IF+JE+\acute{J}\acute{E}\right)^2\left(N+\acute{N}\right)\right)$
When $\acute{J}\acute{E}\ll IF+JE$, the incremental update saves non-negligible computational overhead. 
In scenarios with fast time-varying channels, the incremental model updating for node additions works by fitting the limited amount of newly acquired training data using only a few additional nodes within the time granularity of model updates. Hence, the incremental model updating approach for node additions is effective in these scenarios as well.

Define $\mathbf{Y}_{u}^\text{SE}=\left[\mathbf{Y}_{u}^\text{T},\mathbf{Y}_{u}^\text{a,\text{T}}\right]^\text{T}$.
The overall incremental and collaborative training scheme of the BL model for beam prediction is summarized in Algorithm \ref{algorithm_1}. We discuss its communication overhead and computational complexity in the following. 
Step $2$ of Algorithm \ref{algorithm_1} needs $\mathcal{O}\left(2\acute{N}N_{W}B\bar{K_{u}}IF\right)+\mathcal{O}\left(\acute{N}IFJE\right)$ multiplications. Step $3$ needs $\mathcal{O}\left((N+\acute{N})IF(JE+\acute{J}\acute{E})\right)$ multiplications. As mentioned before, Step $4$ and $5$ need $\mathcal{O}\left(\acute{N}^3\right)+\mathcal{O}\left(\acute{N}^2\left(IF+JE\right)\right)+\mathcal{O}\left(\acute{N}(IF+JE)^2\right)$ and $\mathcal{O}\left(\left(\acute{J}\acute{E}\right)^3\right)+\mathcal{O}\left(\left(\acute{J}\acute{E}\right)^2\left(IF+JE\right)\right)+\mathcal{O}\left(\left(\acute{J}\acute{E}\right)\left(IF+JE\right)^2\right)+\mathcal{O}\left(\left(\acute{J}\acute{E}\right)^2\left(N+\acute{N}\right)\right)+\mathcal{O}\left(\left(\acute{J}\acute{E}\right)\left(N+\acute{N}\right)\left(IF+JE\right)\right)$ multiplications, respectively. Step $6-12$ needs $\mathcal{O}\left((IF+JE+\acute{J}\acute{E})(N+\acute{N})BM\right)+t_\text{max}\mathcal{O}\left((IF+JE+\acute{J}\acute{E})^2BM\right)$ multiplications.
The overall computational complexity per user can be determined by performing a simple summation.
If users adopt D2D communication, the communication overhead per user resulting from Step $9$ is $2t_\text{max}(IF+JE+\acute{J}\acute{E})BM(U-1)$. On the other hand, if the BS facilitates information sharing, the communication overhead per user is $\frac{2t_\text{max}(IF+JE+\acute{J}\acute{E})BM(U+1)}{U}$.
\begin{algorithm}
	\label{algorithm_1}  
	\caption{User-side Incremental and Collaborative Modeling Training} 
	\LinesNumbered %要求显示行号
	\KwIn{$\rho$, $\lambda$, $\mathbf{X}_{u}^\text{a}$, $\mathbf{Y}_{u}^\text{a}$, $\mathbf{C}_u$, $\mathbf{W}_{u,e_{i}}$, $\boldsymbol{\beta}_{u,e_{i}}$, $\mathbf{W}_{u,h_{j}}$, $\boldsymbol{\beta}_{u,h_{j}}$, $i=1,..,I, j=1,...J, u\in\mathbb{U}$}
	\KwOut{$\mathbf{W}_{u}^\text{new},u\in\mathbb{U}$} %输出
	
	\textbf{Initialization:} $\mathbf{W}_{u,h_{j}}$, $\boldsymbol{\beta}_{u,h_{j}}$, $j=J+1,...,J+\acute{J}$, $\mathbf{O}_{u}\left(0\right)$, $\mathbf{W}_{0}\left(0\right)\in \mathbb{R}^{(IF+JE+\acute{J}\acute{E})\times BM}$,$u\in\mathbb{U}$\;
	Use Eq. \eqref{eq_18_zc1} and Eq. \eqref{eq_19_zc1} to calculate $\mathbf{A}_{u}^\text{a}$ and construct $\mathbf{A}_{u}^\text{S}$, $u\in\mathbb{U}$\;
	Use Eq. \eqref{eq_18_zc1} and Eq. \eqref{eq_19_zc1} to calculate $\mathbf{H}_u^\text{a}$ and construct $\mathbf{A}_{u}^\text{SE}$, $u\in\mathbb{U}$\;
	Use Eq. \eqref{eq_30_zc1} to calculate $\mathbf{C}_{u}^\text{S}$, $u\in\mathbb{U}$\;
	Use Eq. \eqref{eq_32_zc1} to calculate $\mathbf{C}_{u}^{\text{SE}}$, $u\in\mathbb{U}$\;
	\For{$t = 1 \rightarrow t_{\max}$}{
		\For{$u = 1 \rightarrow U$}{
			$\mathbf{W}_{u}\left(t\right)=\mathbf{C}_{u}^\text{SE}\left[\mathbf{A}_{u}^\text{SE,\text{T}}\mathbf{Y}_{u}^\text{SE}-\rho\left(\mathbf{O}_{u}\left(t-1\right)-\mathbf{W}_{0}\left(t-1\right)\right)\right]$\;
		   $\overline{\mathbf{W}}(t)=\frac{1}{U}\sum_{\bar{u}=1}^{U}\mathbf{W}_{\bar{u}}(t)$, 
		   $\overline{\mathbf{O}}(t-1)=\frac{1}{U}\sum_{\bar{u}=1}^{U}\mathbf{O}_{\bar{u}}(t-1)$\;
			Use Eq. \eqref{eqn_opt_336} and Eq. \eqref{eqn_opt_337} to calculate $\mathbf{W}_{0}\left(t\right)$ and $\mathbf{O}(t)$\;
		}
	}
	$\mathbf{W}_{u}^\text{new}=\mathbf{W}_{u}\left(t_{\text{max}}\right),u\in\mathbb{U}$\;
\end{algorithm}

{With notably increasing user mobility, e.g, when users traverse considerable distances during the model update interval, leading to a significant lack of channel spatial similarity between the regions in front and behind, the effectiveness of incremental model updating may be compromised, and more reasonable model-update mechanisms should be studied in our future work.}

\section{BS-Side Incremental Collaborative BL-aided BA Design}
\begin{figure}[!h]
%是可选项 h表示的是here在这里插入,t表示的是在页面的顶部插入
\centering
\includegraphics[scale=0.32]{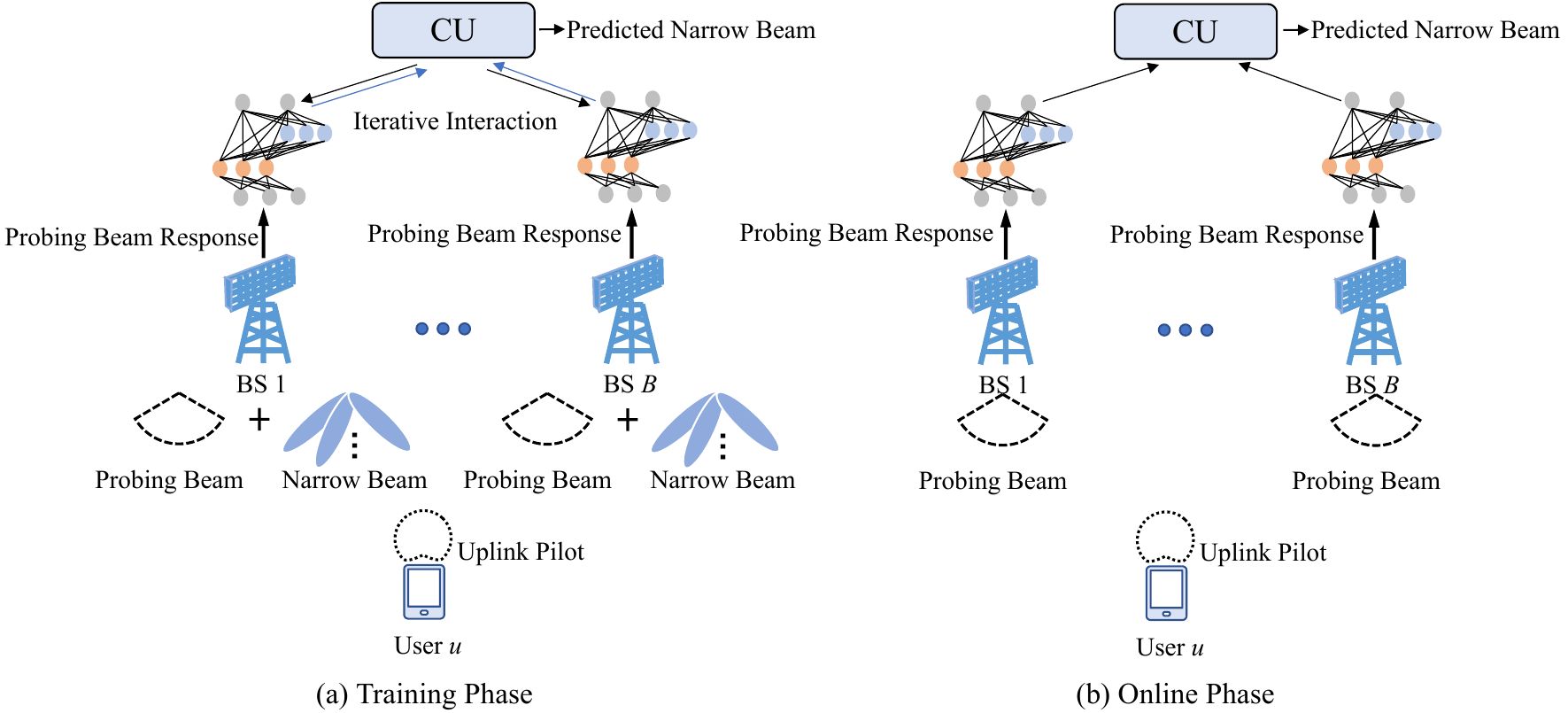}
\caption{{Execution flow of the BL-based beam prediction at the BS side.}}
\label{fig:4}
\end{figure}

In this section, we propose a BL-based incremental collaborative uplink BA method for mmWave cell-free MIMO downlink systems with uplink-downlink channel reciprocity. 
In the offline phase, multiple BSs collect the uplink probing-beam and narrow-beam responses to train the BL models, which are then utilized in the online phase to predict the optimal downlink narrow beams using only the probing beam training.
Fig. \ref{fig:4} illustrates the execution flow.
Note that the features for narrow beam prediction are constituted by the probing-beam responses of multiple BSs.
Nonetheless, directly aggregating beam responses from all BSs to the CU through the fronthaul link results in significant overhead.
To address this issue, we propose a distributed model training scheme inspired by the principles of vertical federated learning, incorporating an incremental update version.

\subsection{Uplink Beam Training}
The receiving signal at BS $b\in \mathbb{B}$ of pilots from user $u\in\mathbb{U}$ on subcarrier $k\in\mathbb{K}_u$ for the $i$-th beam can be represented as
$
{y}_{b,u,k}^{i}=\mathbf{g}_{b}^{i,\text{H}}\mathbf{h}_{b,u,k}{s}_{u,k}^\text{pilot}+\mathbf{g}_{b}^{i,\text{H}}\mathbf{v}_{b,u,k},
$
where ${s}_{u,k}^\text{pilot}=\sqrt{P_{u,k}^\text{tr}}$ with $P_{u,k}^\text{tr}$ denoting the uplink training power. $\mathbf{v}_{b,u,k}\sim\mathcal{CN}\left(0,{\sigma}^{2}\mathbf{I}_{M}\right)$ is the received noise vector. After the pilot matching, BS $b$ can obtain the estimate of each beam response $\mathbf{g}_{b,u}^{i,\text{H}}\mathbf{h}_{b,u,k}$ as 
$
{r}_{b,u,k}^{i}=\mathbf{g}_{b,u}^{i,\text{H}}\mathbf{h}_{b,u,k}+\frac{\mathbf{g}_{b,u}^{i,\text{H}}\mathbf{v}_{b,u,k}}{\sqrt{P_{u,k}^{tr}}}.
$
{According to the previous definition of subcarrier group, the beam response of subcarrier group $\bar{k}$ is $\hat{\mathit{r}}_{b,u,\bar{k}}^{i}={\sum_{k\in {\mathbb{K}}_{u}^{(\bar{k})}}{\mathit{r}}_{b,u,{k}}^{i}}/{\left|{\mathbb{K}}_{u}^{(\bar{k})}\right|}$.}

\subsection{BS-side Collaborative BL Modeling}
In the offline phase, BS $b\in \mathbb{B}$ collects $N$ samples for user $u\in \mathbb{U}$, i.e., the narrow-beam rate {
$\left\{ \mathit{\hat{c}}_{b,u,i}^{\left(n\right)}=\sum\limits_{\bar{k}\in\bar{\mathbb{K}}_{u}}\left|{\mathbb{K}}_{u}^{(\bar{k})}\right|\log_{2}\left(1+\frac{\bar{P}_{u,\bar{k}}}{\sigma^{2}}\left|\hat{r}_{b,u,\bar{k}}^{i,(n)}\right|^{2}\right)\left|\right.i\in\mathbb{M}\right\}$ with $\bar{P}_{u,\bar{k}}={\sum_{k\in {\mathbb{K}}_{u}^{(\bar{k})}}{P}_{u,{k}}}/{\left|{\mathbb{K}}_{u}^{(\bar{k})}\right|}$} and the {probing}-beam responses {$\left\{ \mathit{\hat{r}}_{b,u,\bar{k}}^{i,\left(n\right)}\left|\right.i= -N_{\mathbf{W}}+1,...,0, \bar{k}\in\bar{\mathbb{K}}_{u}\right\}$} for $n=1,...,N$.
For the centralized BL modeling for user $u\in \mathbb{U}$, the $n$-th input is
{$\mathbf{x}_{u}^{\left(n\right)}=\left[\mathbf{x}_{1,u}^{\left(n\right),{\text{T}}},\mathbf{x}_{2,u}^{\left(n\right),{\text{T}}},\ldots,\mathbf{x}_{B,u}^{\left(n\right),{\text{T}}}\right]^{\text{T}}\in\mathbb{R}^{2N_{\mathbf{W}}B\bar{K}_{u}\times1}$ with $\mathbf{x}_{b,u}^{\left(n\right)}=\left[\mathbf{x}_{-N_{\mathbf{W}}+1,b,u}^{\left(n\right),\text{T}},\mathbf{x}_{-N_{\mathbf{W}}+2,b,u}^{\left(n\right),\text{T}},\ldots,\mathbf{x}_{0,b,u}^{\left(n\right),\text{T}}\right]^{\text{T}}\in\mathbb{R}^{2N_{\mathbf{W}}\bar{K}_{u}\times1}$ for $b\in\mathbb{B}$ 
and $
\mathbf{x}_{i,b,u}^{\left(n\right)}=\left[\left|\hat{\mathit{r}}_{b,u,\bar{k}_{u,1}}^{i,\left(n\right)}\right|,\angle\hat{\mathit{r}}_{b,u,\bar{k}_{u,1}}^{i,\left(n\right)},\ldots,\left|\hat{\mathit{r}}_{b,u,\bar{k}_{u,\bar{K}_{u}}}^{i,\left(n\right)}\right|,\angle\hat{\mathit{r}}_{b,u,\bar{k}_{u,\bar{K}_{u}}}^{i,\left(n\right)}\right]^{\text{T}}\in\mathbb{R}^{2\bar{K}_{u}\times1}.
$}
Using the same one-hot coding for $\hat{{\mathbf{{c}}}}_{b,u}^{\left(n\right)}=\left[\hat{{c}}_{b,u,1}^{\left(n\right)},\ldots,\hat{{c}}_{b,u,M}^{\left(n\right)}\right]^{\text{T}}$ as in Eq. \eqref{Eq_15_zc1}, the $n$-th label for learning is denoted as {${\mathbf{y}}_{u}^{\left(n\right)}=\left[{\mathbf{y}}_{1,u}^{\left(n\right),\text{T}},...,{\mathbf{y}}_{B,u}^{\left(n\right),\text{T}}\right]^\text{T}\in\mathbb{R}^{BM\times1}$ with ${\mathbf{y}}_{b,u}^{\left(n\right)} = \bar{{\mathbf{c}}}_{b,u}^{\left(n\right)}$.}
Then, the training set consists of 
${\mathbf{X}}_{u}=\left[{\mathbf{x}}_{u}^{\left(1\right)},\ldots,{\mathbf{x}}_{u}^{\left(N\right)}\right]^{\text{T}}\in\mathbb{R}^{{N\times2N_{\mathbf{W}}B\bar{K}_{u}}}$ and 
${\mathbf{Y}}_{u}=\left[{\mathbf{y}}_{u}^{\left(1\right)},\ldots,{\mathbf{y}}_{u}^{\left(N\right)}\right]^{\text{T}}\in\mathbb{R}^{N\times BM}$.
In addition, the feature and enhancement nodes ${\mathbf{A}}_u$ and the optimization of model weight ${\mathbf{W}}_u$ are mathematically equivalent to those in Eq. \eqref{eq_18_zc1}, Eq. \eqref{eq_19_zc1} and \eqref{eqn_opt_320}. And the idea of using data from multiple users to train a shared model in Eq. \eqref{eqn_opt_324} still applies. To achieve overhead reduction, a distributed executable training scheme is proposed in the following.

First, each BS $b\in\mathbb{B}$ conducts the mapping from its local {probing}-beam response ${\mathbf{X}}_{b,u}=\left[{\mathbf{{x}}}_{b,u}^{\left(1\right)},...,{\mathbf{{x}}}_{b,u}^{\left(N\right)}\right]^{\text{T}}\in\mathbb{R}^{N\times {N_{\mathbf{W}}\bar{K}_{u}}}$ to $I$ groups of feature nodes ${\mathbf{Z}}_{b,u,i}\in\mathbb{R}^{N\times {F}}$ and ${J}$ groups of enhancement nodes ${\mathbf{H}}_{b,u,j}\in\mathbb{R}^{N\times {E}}$, i.e, 
\setlength\abovedisplayskip{1pt}\setlength\belowdisplayskip{1pt}
\begin{align}
{\mathbf{Z}}_{b,u,i}&=\mathit{\phi}\left({\mathbf{X}}_{b,u}{\mathbf{W}}_{b,u,e_{i}}+\mathbf{1}_{N\times1}{\boldsymbol{\beta}}_{b,u,e_{i}}\right),i=1,\ldots,{I}, \label{eqn_Zbui} \\
{\mathbf{H}}_{b,u,j}&=\mathit{\xi}\left({\mathbf{Z}}_{b,u}{\mathbf{W}}_{b,u,h_{j}}+\mathbf{1}_{N\times1}{\boldsymbol{\beta}}_{b,u,h_{j}}\right),j=1,\ldots,{J}, \label{eqn_Hbuj}
\end{align}
where ${\mathbf{W}}_{b,u,e_{i}}$, ${\mathbf{W}}_{b,u,h_{j}}$, ${\boldsymbol{\beta}}_{b,u,e_{i}}$ and ${\boldsymbol{\beta}}_{b,u,h_{j}}$ are the connection weights and bias vectors, which are usually generated randomly and {not trainable}. 
${\mathbf{Z}}_{b,u}=\left[{\mathbf{Z}}_{b,u,1},{\mathbf{Z}}_{b,u,2},\ldots,{\mathbf{Z}}_{b,u,{I}}\right]$ is the cascade matrix of ${I}$ groups of feature nodes. Define ${\mathbf{H}}_{b,u}=\left[{\mathbf{H}}_{b,u,1},{\mathbf{H}}_{b,u,2},\ldots,{\mathbf{H}}_{b,u,{J}}\right]$ as the cascade matrix of ${J}$ groups of enhancement nodes. All these nodes form ${\mathbf{A}}_{b,u}=\left[{\mathbf{Z}}_{b,u}\mid{\mathbf{H}}_{b,u}\right]\in\mathbb{R}^{N\times\left({I}{F}+{J}{E}\right)}$.
%\begin{equation}\label{eqn_Zbui}
%{\mathbf{Z}}_{b,u,i}=\mathit{\phi}\left({\mathbf{X}}_{b,u}{\mathbf{W}}_{b,u,e_{i}}+\mathbf{1}_{N\times1}{\boldsymbol{\beta}}_{b,u,e_{i}}\right),i=1,\ldots,{I},
%\end{equation}
%\begin{equation}\label{eqn_Hbuj}
%{\mathbf{H}}_{b,u,j}=\mathit{\xi}\left({\mathbf{Z}}_{b,u}{\mathbf{W}}_{b,u,h_{j}}+\mathbf{1}_{N\times1}{\boldsymbol{\beta}}_{b,u,h_{j}}\right),j=1,\ldots,{J},
%\end{equation}

Second, the optimization of the model based on feature/enhancement nodes and {labels} from multiple BSs $b\in\mathbb{B}$, i.e., ${\mathbf{A}}_{u}=\left[{\mathbf{A}}_{1,u},...,{\mathbf{A}}_{B,u}\right]$ and {${\mathbf{Y}}_{u}$}, can be formulated as
\setlength\abovedisplayskip{1pt}\setlength\belowdisplayskip{1pt}
\begin{equation}\label{eqn_bsji}
\min_{{\mathbf{W}}_{u}}\frac{1}{2}\left\Vert {\mathbf{A}}_{u}{\mathbf{W}}_{u}-{\mathbf{Y}}_u\right\Vert _\text{F}^{2}+\frac{\lambda}{2}\left\Vert {\mathbf{W}}_{u}\right\Vert_\text{F}^{2},
\end{equation}
where ${\mathbf{W}}_{u}\in\mathbb{R}^{B({I}{F}+{J}{E})\times BM}$ is the affine transformation matrix. {The solution has the same structure as Eq. \eqref{eqn_opt_cen_solu}.}
To reduce the computational stress on the CU and the overhead caused by uploading all features and enhancement nodes to the CU, we transform the above problem into
\begin{equation}
\min_{\left\{{\mathbf{W}}_{b,u},b\in\mathbb{B}\right\}}\frac{1}{2}\left\Vert \sum_{b=1}^{B}{\mathbf{A}}_{b,u}{\mathbf{W}}_{b,u}-{\mathbf{Y}}_u\right\Vert _\text{F}^{2}+\sum_{b=1}^{B}\frac{\lambda}{2}\left\Vert {\mathbf{W}}_{b,u}\right\Vert _\text{F}^{2},
\end{equation}
where ${\mathbf{W}}_{u}=\left[{\mathbf{W}}_{1,u}^\text{T},...,{\mathbf{W}}_{B,u}^\text{T}\right]^\text{T}$ with ${\mathbf{W}}_{b,u}\in \mathbb{R}^{\left({I}{F}+{J}{E}\right)\times BM}$ being the local affine transformation matrix at each BS $b\in\mathbb{B}$. 
According to the distributed model fitting theory in \cite[Section 8.3]{2020Distributed}, the above problem can be expressed as
\setlength\abovedisplayskip{1pt}\setlength\belowdisplayskip{1pt} 
\begin{equation}\label{eqn_opt_418}
\begin{array}{c}
\min\limits_{\left\{{\mathbf{W}}_{b,u},b\in\mathbb{B}\right\}}\frac{1}{2}\left\Vert \sum_{b=1}^{B}\mathbf{V}_{b,u}-{\mathbf{Y}}_u\right\Vert _\text{F}^{2}+\sum_{b=\mathrm{1}}^{B}\frac{\lambda}{2}\left\Vert {\mathbf{W}}_{b,u}\right\Vert _\text{F}^{2},\\
\text{s.t. }{\mathbf{A}}_{b,u}{\mathbf{W}}_{b,u}-\mathbf{V}_{b,u}=\mathbf{0}_{N\times BM},b\in\mathbb{B},
\end{array}
\end{equation} 
where $\mathbf{V}_{b,u}\in\mathbb{R}^{N\times BM}$ is the matrix of introduced auxiliary variables.
%As shown in Fig. \ref{fig:5}, the iterative solving process of problem \eqref{eqn_opt_418} is as follows.
%\begin{figure}[!h]
%\vspace{-0.5cm}
%%是可选项 h表示的是here在这里插入,t表示的是在页面的顶部插入
%\centering
%\includegraphics[scale=0.4]{fig5.pdf}\vspace{-0.5cm}
%\caption{Schematic of iterative interactive weight update at the BS side.}
%\label{fig:5}
%\vspace{-0.5cm}
%\end{figure}
{The iterative solving process of problem \eqref{eqn_opt_418} is conducted as follows.}
\begin{corollary}
	For the $t$-th iteration, {via introducing a dual variable $\mathbf{O}_u \in\mathbb{R}^{N\times BM}$},
	BS $b\in \mathbb{B}$ should conduct
\setlength\abovedisplayskip{1pt}\setlength\belowdisplayskip{1pt}  
\begin{align}
	{\mathbf{W}}_{b,u}\left(t\right)&=\rho\mathbf{Q}_{b,u}^{-1}{\mathbf{A}}_{b,u}^{\text{T}}\left[{\mathbf{A}}_{b,u}{\mathbf{W}}_{b,u}\left(t-1\right)+ \right.\nonumber \\
	&\left.\overline{\mathbf{V}}_u\left(t-1\right)-\overline{{\mathbf{A}}{\mathbf{W}}}_u\left(t-1\right)-\mathbf{O}_u\left(t-1\right)\right], \label{eqn_Wbu}\\
	\overline{\mathbf{V}}_u\left(t\right)&=\frac{1}{B+\rho}\left[\mathbf{Y}_u+\rho\overline{\mathbf{AW}}_u\left(t\right)+\rho\mathbf{O}_u\left(t-1\right)\right], \label{eqn_Vu}\\
	\mathbf{O}_u\left(t\right)&=\mathbf{O}_u\left(t-1\right)+\overline{\mathbf{AW}}_u\left(t\right)-\overline{\mathbf{V}}_u\left(t\right), \label{eqn_Ou}
\end{align}
%\begin{equation}\label{eqn_Wbu}
%{\mathbf{W}}_{b,u}\left(t\right)=\rho\mathbf{Q}_{b,u}^{-1}{\mathbf{A}}_{b,u}^{\text{T}}\left[{\mathbf{A}}_{b,u}{\mathbf{W}}_{b,u}\left(t-1\right)+\overline{\mathbf{V}}_u\left(t-1\right)-\overline{{\mathbf{A}}{\mathbf{W}}}_u\left(t-1\right)-\mathbf{O}_u\left(t-1\right)\right],
%\end{equation}
%\begin{equation}\label{eqn_Vu}
%\overline{\mathbf{V}}_u\left(t\right)=\frac{1}{B+\rho}\left[\mathbf{Y}_u+\rho\overline{\mathbf{AW}}_u\left(t\right)+\rho\mathbf{O}_u\left(t-1\right)\right],
%\end{equation}
%\begin{equation}\label{eqn_Ou}
%\mathbf{O}_u\left(t\right)=\mathbf{O}_u\left(t-1\right)+\overline{\mathbf{AW}}_u\left(t\right)-\overline{\mathbf{V}}_u\left(t\right),
%\end{equation}
where $\mathbf{Q}_{b,u}=\rho\mathbf{A}_{b,u}^{\text{T}}\mathbf{A}_{b,u}+\lambda\mathbf{I}_{IF+JE}$ 
with $\rho>0$ being the penalty coefficient. 
$\overline{\mathbf{AW}}_u(t)=\frac{1}{B}\sum_{b=1}^{B}\mathbf{A}_{b,u}\mathbf{W}_{b,u}(t)$.
\end{corollary}
\begin{IEEEproof}
From \cite[Section 8.3]{boyd2011distributed}, we have
\setlength\abovedisplayskip{1pt}\setlength\belowdisplayskip{1pt}  
\begin{align}
	\mathbf{W}_{b,u}\left(t\right)&=\arg\min_{\mathbf{W}_{b,u}\left(t\right)}\left(\frac{\lambda}{2}\left\Vert \mathbf{W}_{b,u}\left(t\right)\right\Vert _{\mathrm{F}}^{2}+\frac{\rho}{2}\left\Vert \mathbf{A}_{b,u}\mathbf{W}_{b,u}\left(t\right) \right.\right. \nonumber \\
	&\left.\left.-\mathbf{A}_{b,u}\mathbf{W}_{b,u}\left(t-1\right) -\overline{\mathbf{V}}_{u}\left(t-1\right) \right.\right. \nonumber \\
	& +\left.\left.\overline{\mathbf{AW}}_{u}\left(t-1\right)+\mathbf{O}_{u}\left(t-1\right)\right\Vert _{\mathrm{F}}^{2}\right), \label{eqn_yWbu} \\ 
	\overline{\mathbf{V}}_{u}\left(t\right)&=\arg\min_{\overline{\mathbf{V}}_{u}\left(t\right)}\left(\frac{1}{2}\left\Vert B\overline{\mathbf{V}}_{u}\left(t\right)-\mathbf{Y}_{u}\right\Vert _{\mathrm{F}}^{2} \right. \nonumber \\
	&\left. +\frac{B\rho}{2}\left\Vert \overline{\mathbf{AW}}_{u}\left(t\right)-\overline{\mathbf{V}}_{u}\left(t\right)+\mathbf{O}_{u}\left(t-1\right)\right\Vert _{\mathrm{F}}^{2}\right), \label{eqn_yVu}
\end{align} 
and Eq. \eqref{eqn_Ou}. Via using the partial derivative of the object function in Eq. \eqref{eqn_yWbu} with respect to ${\mathbf{W}}_{b,u}\left(t\right)$ and that in Eq. \eqref{eqn_yVu}  
with respect to $\overline{\mathbf{V}}_u\left(t\right)$, Eq. \eqref{eqn_Wbu} and Eq. \eqref{eqn_Vu} can be obtained.
\end{IEEEproof}

The data interactions involved in the iterative process can be implemented based on an architecture of vertical federated learning \cite{chen2020vafl}. Specifically, at the $t$-th iteration, each BS $b$ first updates $\mathbf{W}_{b,u}(t)$ according to Eq. \eqref{eqn_Wbu} and calculates 
$\mathbf{A}_{b,u}\mathbf{W}_{b,u}(t)\in\mathbb{R}^{N\times BM}$. Then, each BS $b$ sends $\mathbf{A}_{b,u}\mathbf{W}_{b,u}$ to the CU via the fronthaul link. After collecting $\mathbf{A}_{b,u}\mathbf{W}_{b,u}$'s from all BSs, the CU calculates $\overline{\mathbf{AW}}_u(t)$. In addition, $\overline{\mathbf{V}}_u\left(t\right)$ and $\mathbf{O}_u\left(t\right)$ can be updated according to Eq. \eqref{eqn_Vu} and Eq. \eqref{eqn_Ou}, respectively. Finally, updated $\overline{\mathbf{AW}}_u(t)$, $\overline{\mathbf{V}}_u\left(t\right)$ and $\mathbf{O}_u\left(t\right)$ are sent to each BS via the fronthual link. This completes one iteration.  

\begin{algorithm}
	\label{algorithm_2}
  	\caption{BS-Side Collaborative BL Model Training} 
  	\LinesNumbered %要求显示行号
 
	\KwIn{$\rho$, $\lambda$, $\mathbf{X}_{b,u}$, $\mathbf{Y}_{b,u}$, $\mathbf{W}_{b,u,e_{i}}$, $\boldsymbol{\beta}_{b,u,e_{i}}$, $\mathbf{W}_{b,u,h_{j}}$, $\boldsymbol{\beta}_{b,u,h_{j}}$, $i=1,..,I, j=1,...,J, b\in\mathbb{B}$}
	\KwOut{$\mathbf{W}_{b,u},b\in\mathbb{B}$}
	\textbf{Initialization:} $\overline{\mathbf{V}}_u\left(0\right)$, $\mathbf{O}_u\left(0\right)$, $\overline{\mathbf{AW}}_u(0)$, ${\mathbf{W}}_{b,u}\left(0\right), b\in\mathbb{B}$\;
	Use Eq. \eqref{eqn_Zbui} and Eq. \eqref{eqn_Hbuj} to calculate $\mathbf{A}_{b,u}$ and $\mathbf{Q}_{b,u}^{-1}$, $b\in\mathbb{B}$\;
	\For{$t = 1 \rightarrow t_{\max}$}{
		\For{$b = 1 \rightarrow B$}{
			Use Eq. \eqref{eqn_Wbu} to update $\mathbf{W}_{b,u}\left(t\right)$ and calculate $\mathbf{A}_{b,u}\mathbf{W}_{b,u}(t)$\;
			Use the MVS method to upload $\mathbf{A}_{b,u}\mathbf{W}_{b,u}(t)$ to the CU\;
		}
		The CU calculates $\overline{\mathbf{AW}}_u(t)$ and updates $\overline{\mathbf{V}}_u\left(t\right)$ via Eq. \eqref{eqn_Vu} and $\mathbf{O}_u\left(t\right)$ via Eq. \eqref{eqn_Ou}\;
		The CU uses the MVS method to deliver $\overline{\mathbf{AW}}_u(t)$, $\overline{\mathbf{V}}_u\left(t\right)$, $\mathbf{O}_u\left(t\right)$ to each BS\;
	}
	$\mathbf{W}_{b,u}=\mathbf{W}_{b,u}\left(t_{\text{max}}\right),b\in\mathbb{B}$\;
\end{algorithm}
It is worth noting that sending $\mathbf{A}_{b,u}\mathbf{W}_{b,u}(t)$ from each BS $b$ to the CU and sending $\overline{\mathbf{AW}}_u(t)$, $\overline{\mathbf{V}}_u\left(t\right)$ and $\mathbf{O}_u\left(t\right)$ from the CU to each BS $b$ involves the communication overhead of $4NBM$ parameters for each fronthaul link. 
To reduce this overhead, one approach is to exploit possible parameter sparsity. 
Specifically, rather than transferring all parameters, only a small portion of the most significant parameters are transmitted.
Based on the Top-$k$ Sparsification gradient compression method \cite{shi2019distributed}, we introduce a maximum value-based sparsification (MVS) method in which for each BS $b$ and each sample $n=1,...,N$, only $N_{b}$ elements with the largest absolute values and their corresponding indices, are transferred from the $n$-th row of the above four matrices in each iteration.
%That is, instead of uploading and downloading all parameters, only a small portion of the most important parameters are transferred. 
%Referring to the Top-$k$ Sparsification gradient compression method \cite{shi2019distributed}, we propose a maximum value-based sparsification (MVS) method in which for each BS $b$ and each sample $n=1,...,N$, only $N_{b}$ elements of the $n$-th row of the above four matrices with largest absolute value and their indices are transferred for each iteration. 
Therefore, the number of parameters to be transferred can be reduced to $8NN_b$.
Algorithm \ref{algorithm_2} outlines the procedure of BS-side collaborative BL training. 
PLease refer to Corollary \ref{corollary2_zc} and \ref{corollary3_zc} for the incremental implementation of $\mathbf{Q}_{b,u}^{-1}=\left(\rho\mathbf{A}_{b,u}^{\text{T}}\mathbf{A}_{b,u}+\lambda\mathbf{I}_{IF+JE}\right)^{-1}$.

In the BS-side collaborative BL-aided BA design, for each BS $b$ to acquire its BL model ${\mathbf{W}}_{b,u}$, the cost is analyzed as follows. 
The computational complexity per BS is 
$
\mathcal{O}\left(2NN_{W}\bar{K_{u}}IF\right)+\mathcal{O}\left(NIFJE\right)+\mathcal{O}\left((IF+JE)^{2}N\right)
$
$+\mathcal{O}\left((IF+JE)^{3}\right)+t_\text{max}\mathcal{O}\left((IF+JE)NBM\right)
$, where the first two terms results from Eq. \eqref{eqn_Zbui}-\eqref{eqn_Hbuj}, and the last three terms comes from Eq. \eqref{eqn_Wbu}.
The communication overhead per BS is $8t_\text{max}NN_{b} + NM$ where the $2$nd term comes from aggregating labels from all BSs to form $\mathbf{Y}_{u}$. 

In contrast, for the centralized training based on data aggregation, i.e., Eq. \eqref{eqn_Zbui}, \eqref{eqn_Hbuj}, \eqref{eqn_bsji}, the per-BS overhead is about $N(IF+JE+M)$, and the per-BS complexity is $\mathcal{O}\left(2NN_{W}\bar{K_{u}}IF\right)+\mathcal{O}\left(NIFJE\right)+\mathcal{O}\left((IF+JE)^{2}NB\right)
+\mathcal{O}\left((IF+JE)^{3}B^2\right)+\mathcal{O}\left((IF+JE)NBM\right)$.
By comparison, we know that if the iteration number $t_\text{max}$ is relatively small, when $IF+JE \gg BM \ge N_b$, the collaborative training significantly saves the communication overhead compared to the centralized training. 
For small $B$ and $t_\text{max}$, the computation complexity of collaborative training and that of centralized training are comparable.  
For large-scale BS cooperation, the amount of complexity savings from the collaborative training depends on $t_\text{max}$.

In the online execution phase, $U$ users first send their pilots for training {probing} beams with time cost {$N_{\mathbf{W}}T_\text{b}$}. Then, BS $b\in\mathbb{B}$ uses local {probing}-beam response from user $u\in\mathbb{U}$, i.e., $\mathbf{x}_{b,u}$, to obtain the local joint feature and enhancement nodes $\mathbf{a}_{b,u}$. In addition, BS $b\in\mathbb{B}$ calculates the local beam prediction $\mathbf{a}_{b,u}^{\text{T}}\mathbf{W}_{b,u}$ and uploads it to the CU. Based on the integrated beam prediction $\hat{\mathbf{y}}_{u}^{\text{T}} = \sum_{b=1}^{B}\mathbf{a}_{b,u}^{\text{T}}\mathbf{W}_{b,u}$, the CU determines the beam index of BS $b\in\mathbb{B}$ for user $u\in\mathbb{U}$ as $I^{\star}_{b,u}=\arg \max_{i\in{\mathbb{M}}}\{\hat{{y}}_{(b-1)M+i,u}\}$ and sends it to BS $b\in\mathbb{B}$. The effective downlink rate is 
\setlength\abovedisplayskip{1pt}\setlength\belowdisplayskip{1pt} 
\begin{equation}
	\begin{aligned}
		R_{u}^{\text {eff }}=&\left(1-\frac{{(N_{\mathbf{W}}+1)}T_\text{b}}{T}\right) \frac{B_{w}}{K} \times \\
		&\hspace{0.5cm}\sum_{k \in \mathbb{K}_{u}} \log _{2}\left(1+\frac{P_{u,k}}{\sigma^{2}} \sum_{b=1}^{B}\left|\mathbf{h}_{b, u, k}^\text{H} \mathbf{f}_{I^{\star}_{b,u}}\right|^{2}\right),
	\end{aligned}
\end{equation}
for user $u$ where ${(N_{\mathbf{W}}+1)}{T_\text{b}}$ is the time spent by the beam training for {probing} beams and {predicted} narrow beams for transmission.
Note that multiple users and BSs can conduct the beam training simultaneously in uplink training based on the OFDMA mode. {The computational complexity and communication overhead per BS in the online execution phase are $\mathcal{O}\left(2{N_{\mathbf{W}}{\bar{K}_{u}}IF}\right) + \mathcal{O}\left(IFJE\right)+\mathcal{O}\left((IF+JE)BM\right)$ and $BM$. }

\section{Simulation and Discussion}
In this section, we evaluate the performance of the proposed user-side and BS-side incremental collaborative BA schemes. 
For system setup and channel generation, we adopt the system and channel model in Section \ref{sec2}.
We use the DeepMIMO channel dataset \cite{Alkhateeb2019DeepMIMO}, created by the commercial ray-tracing simulator Wireless InSite \cite{eichenlaub2008fidelity}, to ensure the reasonableness of relevant parameter settings, e.g., path gain, angle, and delay, etc. This dataset is widely used in mmWave research \cite{va2017inverse}, and has been verified with channel measurements \cite{li2015validation}.
\begin{figure}[ht!]
\centering
\includegraphics[height=4.5cm, width=8cm]{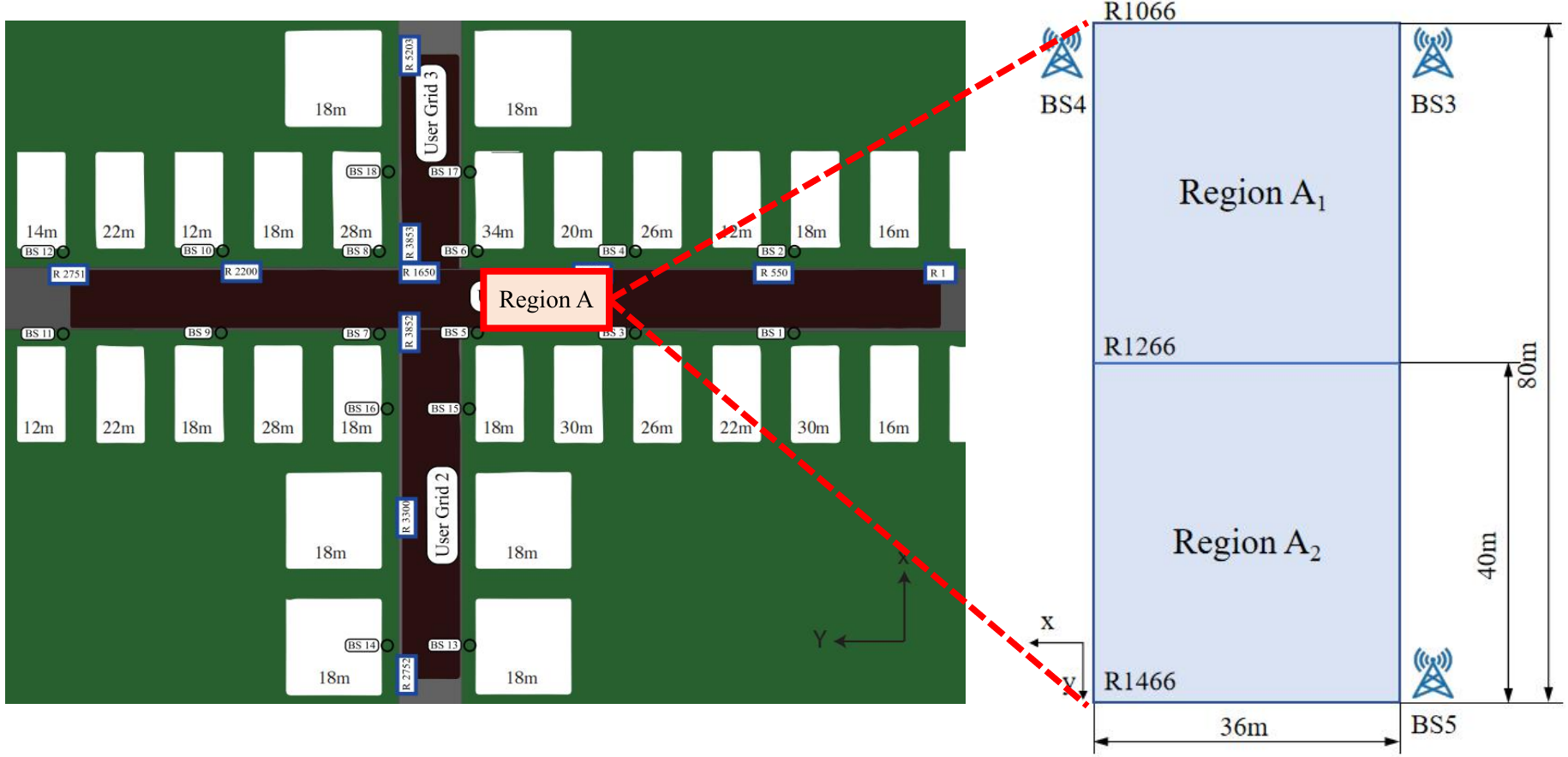}
\caption{{The mmWave cell-free MIMO scenario.}}
\label{fig:6_1}
\end{figure}
\begin{figure}[ht!]
\centering
\includegraphics[height=4.5cm, width=7cm]{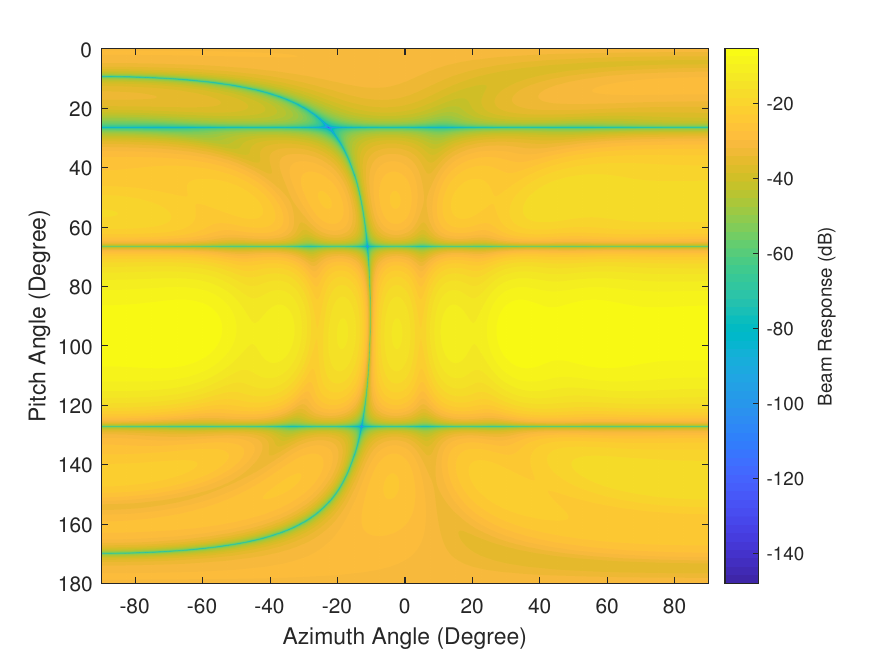}
\caption{{The wide-beam response of BS $3$.}}
\label{fig:6_2} 
\end{figure}
The 'O1' scenario of DeepMIMO dataset is chosen for the following simulations. 
{Fig. \ref{fig:6_1} gives the top view of the 'O1' scenario from which we select three BSs that are numbered $3$, $4$ and $5$, and Region $A$ of $36$m$\times$$80$m in the main street (marked by the red box) including Region $A_1$ with rows R$1066$ to R$1266$ and Region $A_2$ with rows R$1267$ to R$1466$ to build the simulation scenario.
$401\times91=36491$ locations are sampled in Region $A$ at $0.4$m and $0.2$m intervals along the $x$-direction and $y$-direction, respectively.}
{The mobile user setup is similar to that in \cite{alkhateeb2018deep}, i.e., at every beam coherence time $T$, the location of the mobile user is randomly selected from Region $A$ in Fig. \ref{fig:6_1}. Specifically, $29193$ locations randomly selected from Region $A$ form the training dataset while the remaining $7298$ locations form the testing dataset. For scenarios with $U$ users, the training dataset and testing dataset are divided into $U$ parts, which respectively constitute each user's training and testing dataset.}

The system is with carrier frequency $60$ GHz, the bandwidth of $500$ MHz, and $1024$ subcarriers, where three BSs each with an $8\times4$ UPA serves {2 (default value) or {8}} single-antenna users via the OFDMA mode. 
Each user occupies {$64$ subcarriers which are divided into $16$ subcarrier groups}. {To embody the effect of multipath, the number of paths is set to $3$.} 
The transmit power of the BS and the user is set to be {$5$} W and {$200$} mW, respectively.
{The codebook of narrow beams in the training and that of predicted transmission beams are both the standard UPA two-dimensional DFT matrix.}
%The adopted beam training codebook is a standard UPA two-dimensional DFT matrix. 
{Without specific notation, we construct the probing beam for each BS in both the training and inference phase of our proposed BL-based BA scheme by expanding the steering beam of each BS pointing at the center of Region $A$ in Fig. \ref{fig:6_1}, e.g., $\mathbf{a}_{z}\left(96.25\pi/180\right) \otimes \mathbf{a}_{y}\left(55.89\pi/180, 96.25\pi/180\right)$ for BS $3$, according to \cite[Eq. (7)-(9)]{sergeev2017enhanced} with expansion factor $c=0.9$ and $p=2$ (no expansion for narrow vertical coverage).
Fig. \ref{fig:6_2} shows the $2$D response of BS $3$'s multi-antenna probing beam.
}
{We also adopt the omnidirectional beam excited by a single antenna (in Fig. \ref{fig:10_2} and \ref{fig:11_2}) and the above steering beam (in Fig. \ref{fig:11_2}) as the probing beam for performance comparison.  
}
%{The wide-beam orientating the coverage area formed by all antennas \cite[Eq. (7)-(9)]{sergeev2017enhanced} is considered as the probing beam for learning.} {Fig. \ref{fig:6_2} shows the normalized response of the $2$D wide beam of BS $3$ which is formed by the Kronecker product of $z$-directional beam $\mathbf{a}_{z}\left(96.25\pi/180\right)$ (no expansion for narrow vertical coverage) and the $y$-directional beam, i.e., the expanded version of $\mathbf{a}_{y}\left(55.89\pi/180, 96.25\pi/180\right)$ (processed according to \cite[Eq. (7)-(9)]{sergeev2017enhanced} with expansion factor $c=0.9$ and $p=2$).}
{The beam coherence time $T$ is calculated according to \cite[Eq. (8) and (50)]{va2016impact}. Its default value is $96$ ms corresponding to vehicle scenarios with moving speed $v \approx 30$ mph.} $T_\text{b}$ is set to be $0.48$ ms. 
 
We consider seven schemes for the performance comparison:
\begin{enumerate}
    \item \underline{\textbf{ICBL Scheme}}: For our proposed incremental collaborative BL (ICBL) based beam prediction scheme, we set $I=10$, $F=20$, $J=1$. $E$ is  adjusted according to the number of training samples $N$, i.e., {$E=500$ for $N < 1000$, $E=1500$ for $N\ge 1000$}. The feature generation layer and the feature enhancement layer, respectively, adopt the linear and Tansig non-linear activation functions. {The number of iterations $t_\text{max}$ is $10$ and $5$ for UE-side and BS-side collaborative learning, respectively.} In addition, $\rho = 0.1$, {$\lambda = 2^{-3}$} for the user-side learning and {$\lambda = 2^{-9}$} for the BS-side learning. 
    \item {\underline{\textbf{CBL Scheme}}: Compared to the ICBL scheme, the only difference is that the incremental model update is removed}.
    \item \underline{\textbf{FCBL Scheme}}: For the fully centralized BL (FCBL) based beam prediction scheme, samples from {multiple} users and those from $3$ BSs are brought together for the user-side learning and the BS-side learning, respectively. We refer to the related BL parameters in the ICBL scheme for parameter settings.
    \item \underline{\textbf{FDBL Scheme}}: The fully distributed BL (FDBL) based beam prediction scheme requires no inter-user or inter-BS collaboration. That is to say, each BS or user only uses local data to train the beam prediction model. Related BL parameters are the same as those of the ICBL scheme. {The following performance is averaged across models at all BSs or UEs.}
	\item \underline{\textbf{DNN scheme}}: To improve the learning efficiency, the DNN-based regression model for centralized beam prediction in  \cite{chen2021distributed} is modified to a classification model with a Softmax output layer. {Each DNN model for one BS's beam prediction has $2$ hidden layers. The $1$st hidden layer is with $200$ ReLU neurons, and the number of ReLU neurons in the $2$nd layer is the same as that of enhancement nodes in the ICBL scheme}.
	The dropout ratio and learning rate are set to be $0.05$ and $0.001$. The batch size is set to be $100$. 
	The Adam optimizer is used to update the DNN model under the Cross-Entropy Loss Function (CLF). The TensorFlow and Keras libraries are used in the simulation.
	\item {\underline{\textbf{Genie-Aided scheme}}: The Genie-Aided scheme can perform the optimal BA with no training overhead (ideal case).}
	\item {\underline{\textbf{Exhaustive-search}}:  The exhaustive-search-based BA method performs an exhaustive search of the candidate beams in the codebook.}
\end{enumerate}
The performance metric is the effective rate averaged over users and subcarriers for better clarification, i.e., 
\setlength\abovedisplayskip{1pt}\setlength\belowdisplayskip{1pt}
\begin{equation}
	\begin{aligned}
		\text{SE}_{\text{ave}}^{\text {eff }}=&\frac{\left(1-{T_\text{r}}/{T}\right)}{U} \times \\
		&\sum_{u\in \mathbb{U}} \left(\frac{1}{K_u} \sum_{k \in \mathbb{K}_{u}} \log _{2}\left(1+\frac{P_{u,k}}{\sigma^{2}} \sum_{b=1}^{B}\left|\mathbf{h}_{b, u, k}^\text{H} \mathbf{f}_{I^{\star}_{b,u}}\right|^{2}\right)\right).
	\end{aligned}
\end{equation}

\subsection{Beam Prediction for User-Side Learning}

\begin{figure}[htb]
\centering
\includegraphics[scale=0.5]{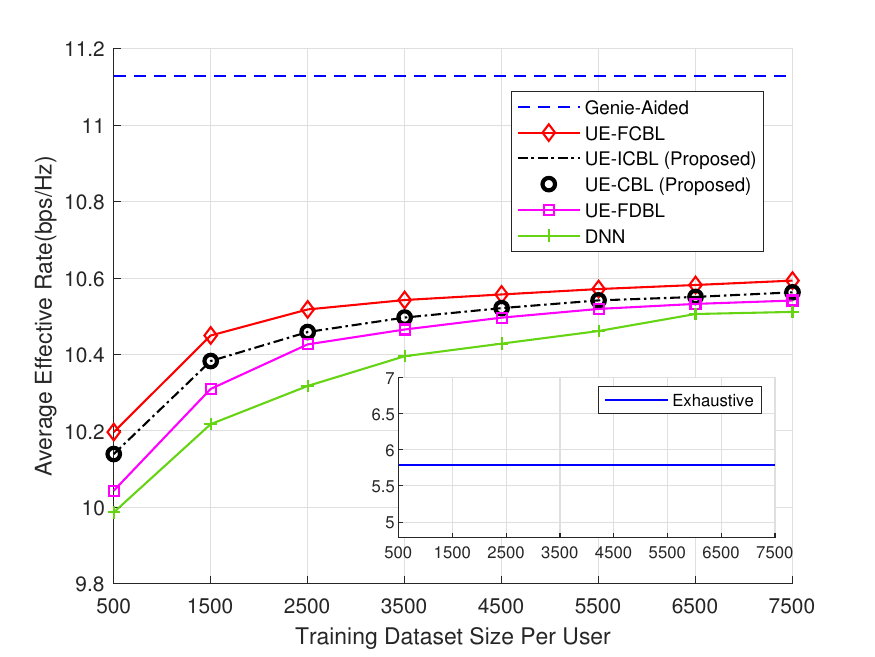} % fig1_left
\caption{{$\text{SE}_{\text{ave}}^{\text {eff }}$ for $2$ users' collaboration.}}
\label{fig:7_1}
\end{figure}

\begin{figure}[htb]
\centering
\includegraphics[scale=0.5]{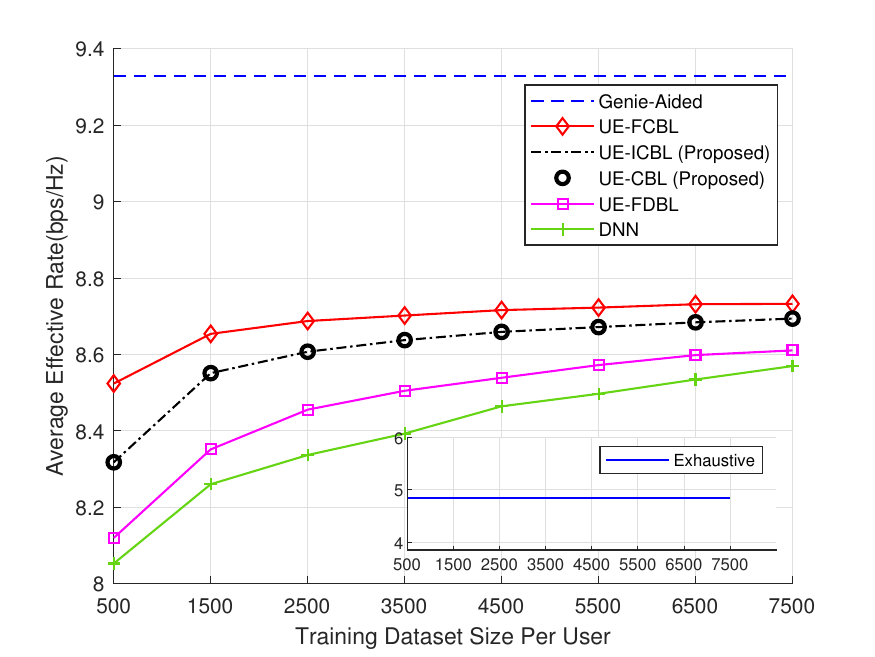} % fig1_medium
\caption{{$\text{SE}_{\text{ave}}^{\text {eff }}$ for $8$ users' collaboration.}}
\label{fig:7_2}
\end{figure}

In Fig. \ref{fig:7_1} and Fig. \ref{fig:7_2}, we compare the average effective rate of seven schemes for $U=2$ and $U=8$, respectively.
{First, the BL-based schemes, i.e., ICBL, CBL, FCBL, and FDBL, perform better than the DNN scheme. But the gap is gradually narrowing with increasing training dataset size. This shows that for the interval with the relatively small size of the training dataset, the BL model with a {broader} structure and untrainable feature/enhancement nodes demonstrates better generalization ability.  
Second, the proposed CBL scheme and ICBL scheme have the same performance, demonstrating the effectiveness of incremental model updating. And their performance is better than that of the FDBL scheme and is close to that of the FCBL scheme. This verifies the necessity and effectiveness of implicit sharing of multi-user datasets via collaborative training.
Third, the collaboration gain, i.e., the gap between ICBL and FDBL, becomes large when the number of collaborative users increases from $2$ to $8$.} 
Finally, compared to the case of uplink training and BS-side learning-based beam prediction, here the performance advantage of learning-based schemes over the exhaustive-search BA scheme is more significant. This is because {for the downlink narrow beam training}, the pilot overhead is proportional to the total antenna number of multiple BSs.

\begin{figure}[ht!]
\centering
\includegraphics[scale=0.5]{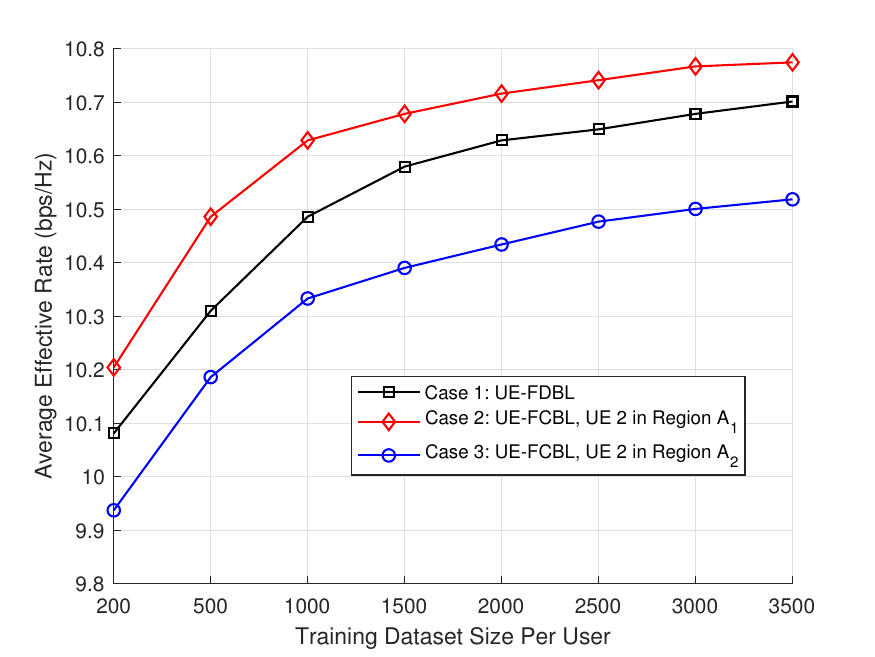} % fig1_medium
\caption{Impact of UE activity area similarity on collaboration.}
\label{fig:8_2}
\end{figure}
\begin{figure}[ht!]
\centering
\includegraphics[scale=0.5]{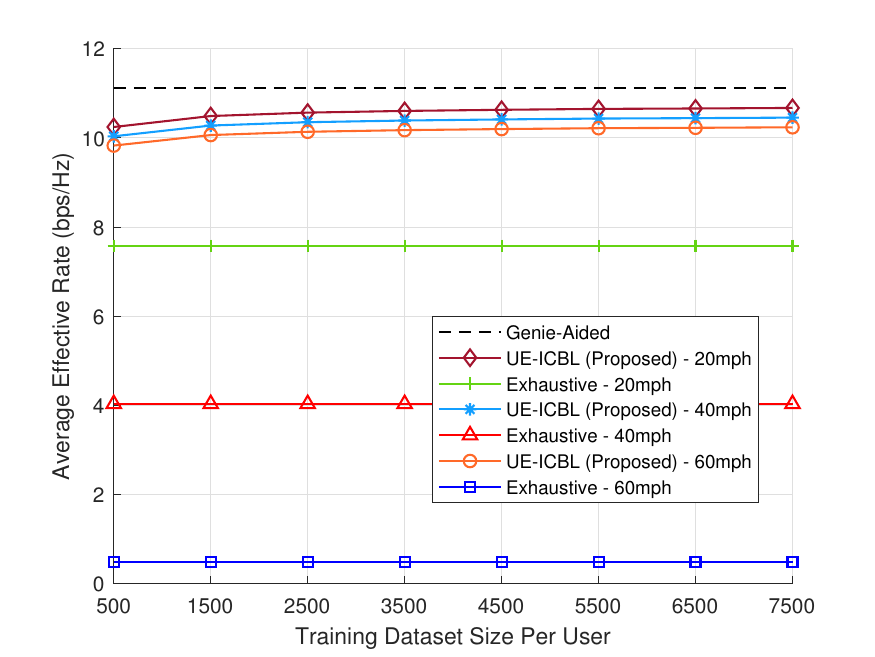}% fig1_medium
\caption{{Average effective rate versus user movement speed.}}
\label{fig:9}
\end{figure}
{For user-side learning, the mismatch between the local data distribution and the global data distribution needs to be considered. We divide Region $A$ into $2$ parts, i.e., Region $A_1$-$A_2$, as shown in Fig. \ref{fig:6_1}, and consider a two-user scenario where the location of UE $1$ is randomly sampled from Region $A_1$ while the location of UE $2$ is randomly sampled either from Region $A_1$ or $A_2$.} The distance between the centers of these two regions is $40$m. {Fig. \ref{fig:8_2} shows the average effective rate of UE $1$ in three cases. Case $1$: UE $1$ only uses local data to perform the FDBL scheme; Case $2$: UE $1$ and UE $2$ both located in Region $A_1$ conduct the FCBL scheme; Case $3$: UE $1$ located in Region $A_1$ and UE $2$ located in Region $A_2$ conduct the FCBL scheme.  
We can see that when two users are located in the same Region $A_1$, the collaboration helps improve UE $1$'s performance.}
 However, when UE $2$ is located in Region $A_2$, the mismatch between two users' channel statistics makes user collaboration degrade UE $1$'s performance.

Fig. \ref{fig:9} shows the effect of user movement speed on the average effective rate of the system. As the user speed increases from $20$ mph to $60$ mph, {the beam coherence time $T$ decreases from $144.54$ ms to $48.18$ ms and the data transmission interval $T-T_\text{r}$ declines accordingly}, resulting in a lower effective rate for both the proposed ICBL scheme and the exhaustive search scheme. However, the impact of speed increase on the performance of the proposed ICBL is significantly much smaller. And it can still achieve an effective rate higher than {$9.8$} bps/Hz in scenarios with high mobility, e.g., $60$ mph.

\subsection{Beam Prediction for BS-Side Learning}

\begin{figure}[ht!]
\centering
\includegraphics[scale=0.5]{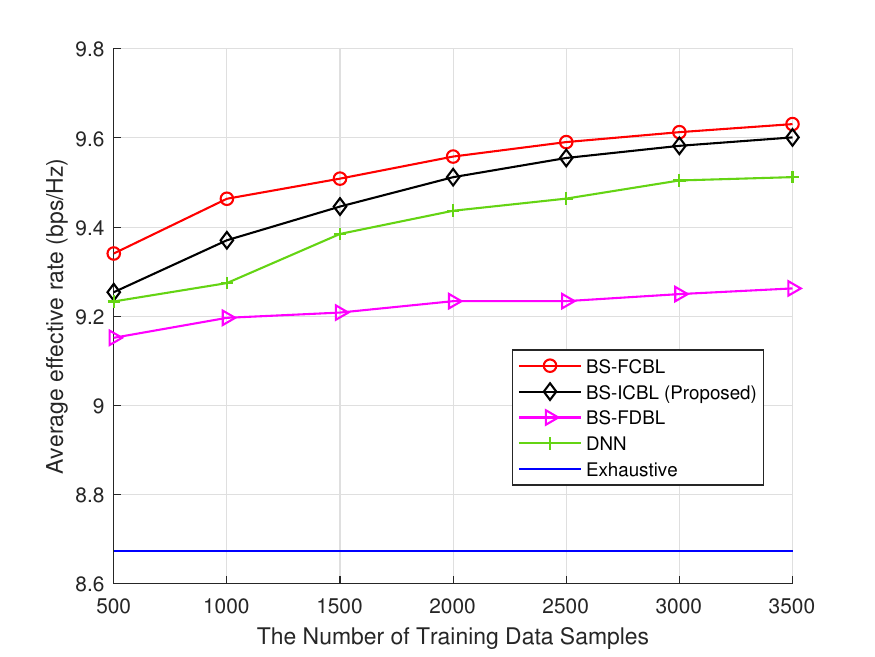} % fig1_left
\caption{{Performance with a multi-antenna wide beam.}}
\label{fig:10_1}
\end{figure}

For the BS-side learning, to focus on the BS cooperation gain, we consider one user to evaluate the BA performance.
As can be seen from Fig. \ref{fig:10_1}, among the BL-aided BA schemes, the FCBL and ICBL schemes perform significantly better than the FDBL scheme.  
This shows that probing-beam measurements from a single BS do not contain enough spatial information to predict the best beam. And feature sharing via BS collaboration is necessary. The DNN scheme based on multiple-BS probing-beam measurements also performs better than the FDBL scheme. The ICBL scheme has better performance than the DNN scheme, showing the advantage of a broad structure for the considered scenario with relatively small training samples.
\begin{figure}[ht!]
\centering
\includegraphics[scale=0.5]{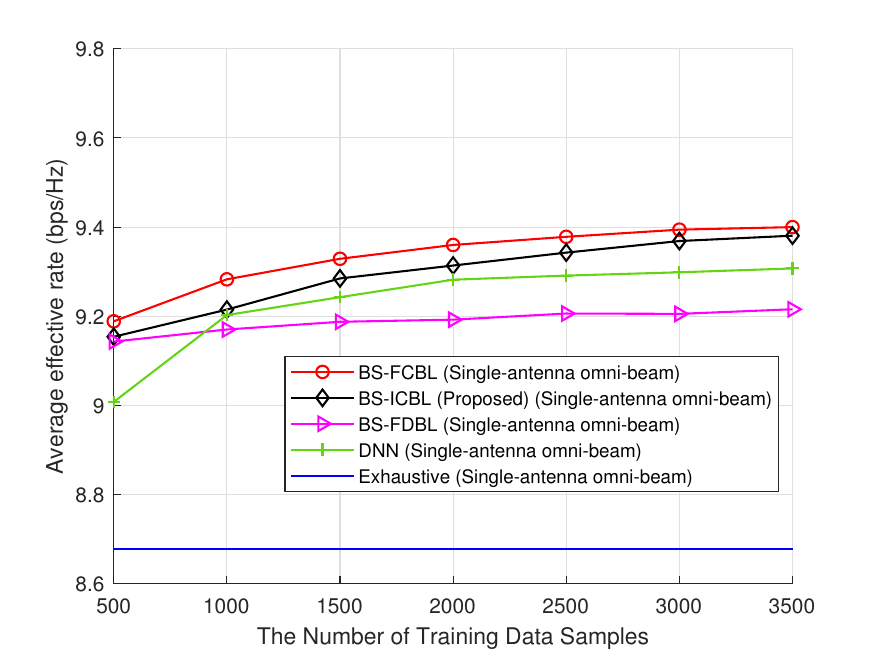} % fig1_medium
\caption{{Performance with a single-antenna omni-beam.}}
\label{fig:10_2}
\end{figure}
 
Since the user transmit power is relatively small, the sensitivity of the probing-beam response-based beam prediction to the user uplink training power is also worth investigating. {Fig. \ref{fig:10_2} shows the performance of BL and DNN-based BA schemes when the BSs use the omnidirectional beam created by a single antenna as the probing beam. Compared to this single-antenna omnidirectional beam, the multi-antenna probing beam helps increase the average effective rate of the ICBL scheme with $3500$ data samples from $9.381$ bps/Hz to $9.613$ bps/Hz, i.e., $2.47\%$ improvement.}
%Compared to the curves in Fig. \ref{fig:10_1}, all learning schemes have different degrees of performance degradation, \textcolor{blue}{e.g., the average effective rate of the ICBL scheme with $3500$ data samples decreases from $9.613$ bps/Hz to $9.381$ bps/Hz, i.e., $2.47\%$ loss.}
%compared to the omnidirectional beam excited by a single antenna, the probing beam formed by multiple antennas aligning to the user's active area helps increase the average effective rate of the ICBL scheme with $3500$ data samples from $9.381$ bps/Hz to $9.613$ bps/Hz, i.e., $2.47\%$ improvement.
This is because the multi-antenna probing beam as shown in Fig. \ref{fig:6_2} has a higher transmitting/receiving gain than the single-antenna omnidirectional beam for the coverage area. And the multi-antenna probing beam draws a better signature of the environment and trains the neural network model more efficiently.

\begin{figure}[ht!]
\centering
\includegraphics[scale=0.5]{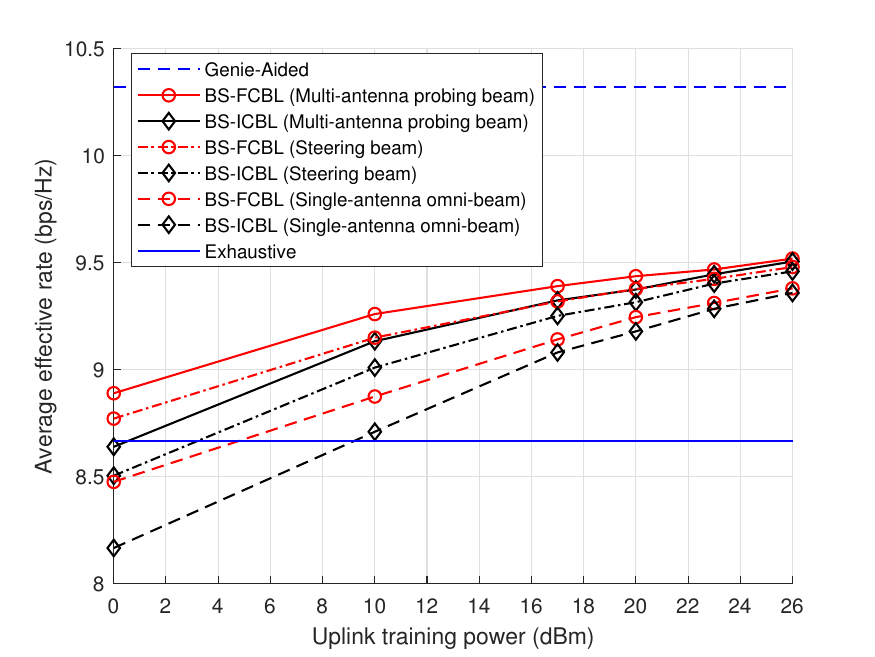} % fig1_medium
\caption{Average effective rate versus uplink training power.}
\label{fig:11_2}
\end{figure}

{Fig. \ref{fig:11_2} shows the average effective rate of the proposed FCBL and ICBL schemes versus different uplink training powers for three different probing beam settings, i.e., 1) the single-antenna omnidirectional beam, 2) the steering beam, 3) the multi-antenna probing beam.}
The size of the training dataset is $1500$.
{With decreasing uplink training power, the advantage of FCBL and ICBL schemes over the exhaustive search tends to decrease. However, it still achieves non-negligible gain over the exhaustive search for relatively low uplink power around $10$ dBm, especially with the multi-antenna probing beam.} 
{In addition, the multi-antenna probing beam results in a higher rate than the other two probing beams, especially with decreasing uplink training power. For example, the rate of FCBL with uplink training power of $0$ dBm is $8.476$ bps/Hz and $8.890$ bps/Hz for the single-antenna omnidirectional beam and the multi-antenna probing beam, respectively, corresponding to a $4.88\%$ increase.
This is because although the steering beam has a higher gain in the direction pointing at the area center, the multi-antenna probing beam takes into account the random uncertainty of the user's location and has a higher average beam gain in the user's active area, compared to the steering beam and the single-antenna omnidirectional beam, thus providing a higher SNR environment for the BS-side learning.} 

\begin{figure}[ht!]
\centering
\includegraphics[scale=0.5]{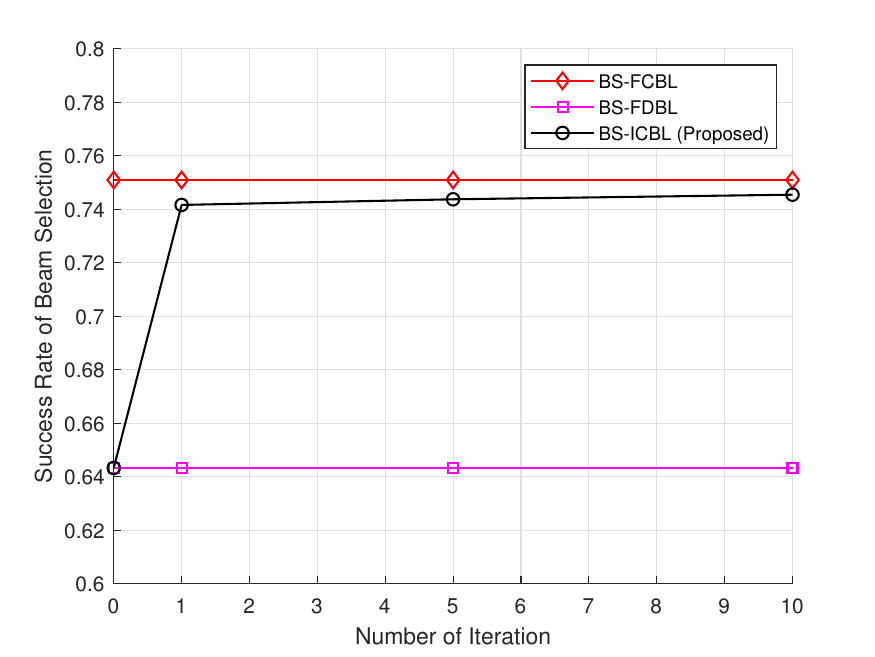} % fig1_left
\caption{{Impact of iteration number on the BA success rate}}
\label{fig:11_1}
\end{figure}
Fig. \ref{fig:11_1} shows the convergence performance of the proposed ICBL scheme, where the number of training samples per BS is {$3000$}. {Successful beam selection means that all BSs make accurate beam prediction}. A very fast convergence can be observed for the proposed schemes.

\begin{figure}[ht!]
\centering
\includegraphics[scale=0.5]{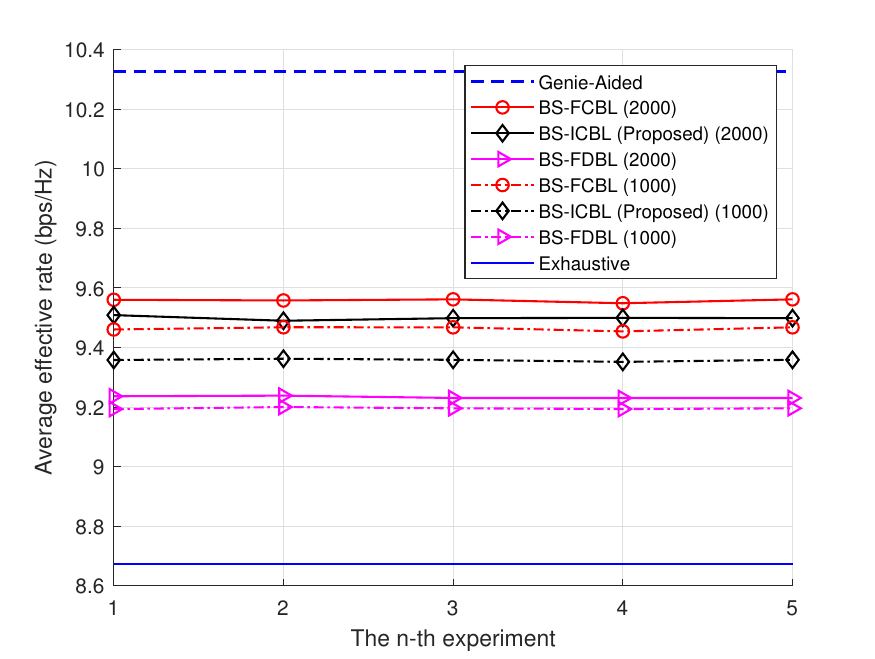} % fig1_left
\caption{{Performance sensitivity to BL network initialization.}}
\label{fig:12_1}
\end{figure}
In Fig. \ref{fig:12_1}, the sensitivity of the proposed schemes with different realizations of untrainable feature/enhancement nodes is demonstrated for the training dataset of $1000$ samples and that of $2000$ samples. It can be seen that the performance of proposed BL-aided BA schemes is barely affected by $5$ different realizations of node weights, showing the robustness of our design.

\begin{figure}[ht!]
\centering
\includegraphics[scale=0.5]{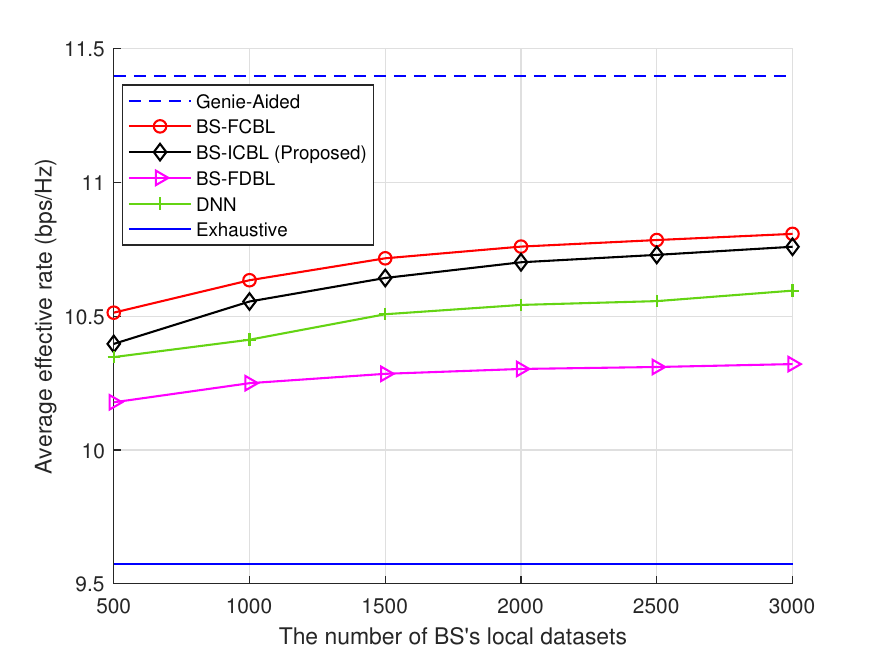} % fig1_medium
\caption{{Performance comparison in $28$ GHz scenarios.}}
\label{fig:12_2}
\end{figure}
In Fig. \ref{fig:12_2}, we study the performance of the proposed schemes in another typical mmWave band, i.e., $28$ GHz. The parameter settings in this scenario are the same as in the $60$ GHz scenario. It can be seen that compared to Fig. \ref{fig:10_1}, the relationship among the performance of all schemes remains the same, but there is an increase in the overall level.  
On the one hand, this proves the effectiveness of the proposed BL-aided BA schemes in the $28$ GHz scenario with a more obvious multipath effect. On the other hand, unlike the multi-cell case where higher path loss in the $60$ GHz scenario reduces inter-cell interference, in the considered joint transmission-based multi-BS scenario, the lower path loss in the $28$ GHz scenario increases the effective rate, due to the higher SNR level in both model training/inference for beam prediction and uplink data decoding with predicted beams.

\section{Conclusion}
To cope with the problem of high BA overhead in mmWave cell-free MIMO downlink systems, the probing beam-based BL-aided BA design was studied in this paper.
For channels without and with uplink-downlink reciprocity, we proposed the user-side and BS-side BL-aided incremental collaborative BA approaches, which respectively realized implicit sharing of multiple user data and multiple BS features via reasonable distributed BL designs. Numerical results verified the applicability of the user-side scheme to scenarios with fast time-varying and/or non-stationary channels and that of the BS-side scheme to scenarios with low fronthaul capacity and CU of less computing power. The advantages of the proposed schemes were also confirmed compared to traditional and DNN-aided BA schemes for these scenarios.

\ifCLASSOPTIONcaptionsoff
  \newpage
\fi

\bibliographystyle{IEEEtran}
% 通过bib调用参考文献
%\bibliography{References}

% 通过bbl调用文献，注意保持更新


\begin{IEEEbiography}[{\includegraphics[width=1in,height=1.25in,clip,keepaspectratio]{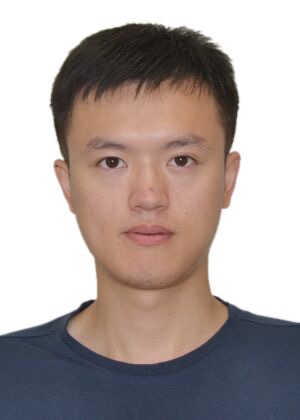}}]{Cheng Zhang}
	(Member, IEEE) received the B.Eng. degree from Sichuan University, Chengdu, China, in June 2009, the M.Sc. degree from the Xi’an Electronic Engineering Research Institute (EERI), Xi’an, China, in May 2012, and the Ph.D. degree from Southeast University (SEU), Nanjing, China, in Dec. 2018. From Nov. 2016 to Nov. 2017, he was a Visiting Student with the University of Alberta, Edmonton, AB, Canada.
	
	From June 2012 to Aug. 2013, he was a Radar Signal Processing Engineer with Xi’an EERI. Since Dec. 2018, he has been with SEU, where he is currently an Associate Professor, and supported by the Zhishan Young Scholar Program of SEU. His current research interests include space-time signal processing and machine learning-aided optimization for B5G/6G wireless communications. He has authored or co-authored more than 40 IEEE journal papers and conference papers. He was the recipient of the excellent Doctoral Dissertation of the China Education Society of Electronics in Dec. 2019, that of Jiangsu Province in Dec. 2020, and the Best Paper Award of 2023 IEEE WCNC.
\end{IEEEbiography}

\begin{IEEEbiography}[{\includegraphics[width=1in,height=1.25in,clip,keepaspectratio]{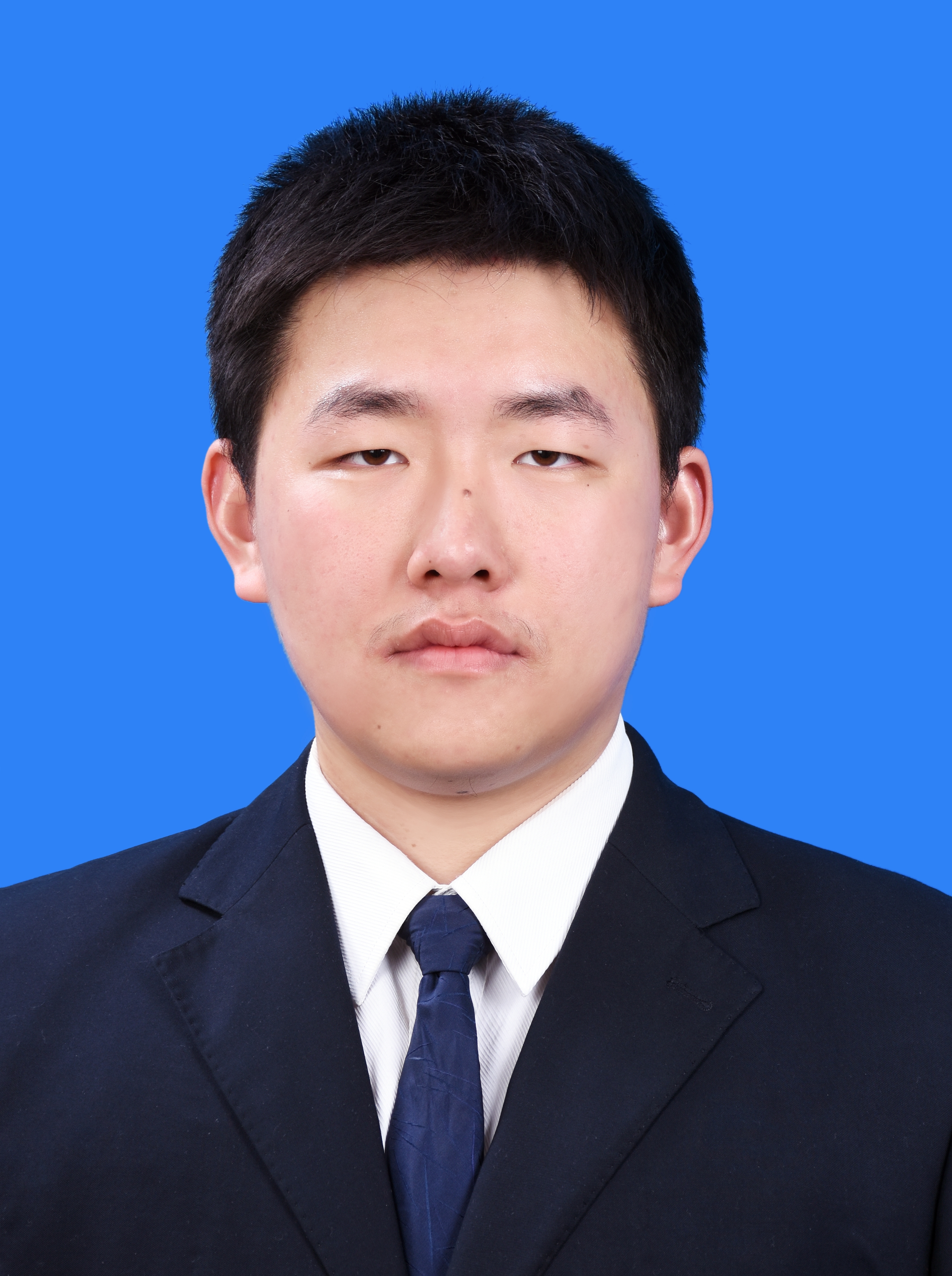}}]{Leming Chen}
	received the B.Eng. degree in electronic information engineering from School of Electronics and Information, Northwestern Polytechnical University, Xi'an, China, in 2020. And he is currently pursuing the M.Sc. degree in information and communication engineering with the School of Information Science and Engineering, Southeast University, Nanjing, China. His research interests mainly focus on intelligent wireless communications.
\end{IEEEbiography}

\begin{IEEEbiography}[{\includegraphics[width=1in,height=1.25in,clip,keepaspectratio]{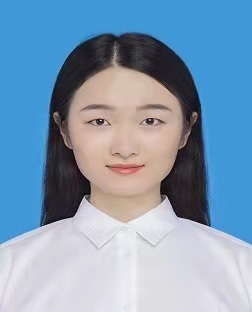}}]{Lujia Zhang}
	received the B.Eng. degree in communication engineering from the School of Communication and Information Engineering, Nanjing University of Posts and Telecommunications, Nanjing, China, in 2019, where she is currently pursuing the M.Sc. degree in information and communication engineering with the School of Information Science and Engineering, Southeast University. Her research interests mainly focus on intelligent wireless communications assisted by machine learning.
\end{IEEEbiography}

\begin{IEEEbiography}[{\includegraphics[width=1in,height=1.25in,clip,keepaspectratio]{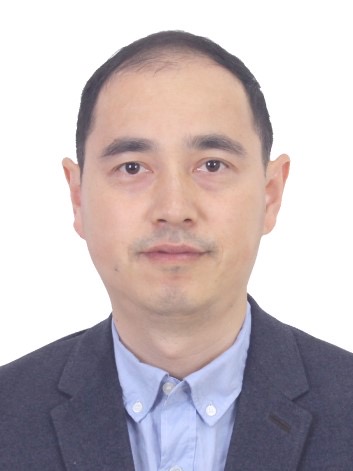}}]{Yongming Huang}
	(M’10-SM’16) received the B.S. and M.S. degrees from Nanjing University, Nanjing, China, in 2000 and 2003, respectively, and the Ph.D. degree in electrical engineering from Southeast University, Nanjing, in 2007.
	
	Since March 2007 he has been a faculty in the School of Information Science and Engineering, Southeast University, China, where he is currently a full professor. He has also been the Director of the Pervasive Communication Research Center, Purple Mountain Laboratories, since 2019. From 2008 to 2009, he was visiting the Signal Processing Lab, Royal Institute of Technology (KTH), Stockholm, Sweden. He has published over 200 peer-reviewed papers, hold over 80 invention patents. His current research interests include intelligent 5G/6G mobile communications and millimeter wave wireless communications. He submitted around 20 technical contributions to IEEE standards, and was awarded a certiﬁcate of appreciation for outstanding contribution to the development of IEEE standard 802.11aj. He served as an Associate Editor for the IEEE Transactions on Signal Processing and a Guest Editor for the IEEE Journal on Selected Areas in Communications. He is currently an Editor-at-Large for the IEEE Open Journal of the Communications Society.
\end{IEEEbiography}

\begin{IEEEbiography}[{\includegraphics[width=1in,height=1.25in,clip,keepaspectratio]{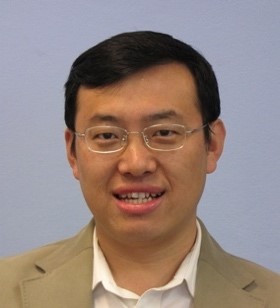}}]{Wei Zhang}
	(Fellow, IEEE) received the Ph.D.
	degree from The Chinese University of Hong Kong
	in 2005. 
	
	He is currently a Professor at the School of Electrical Engineering and Telecommunications, University
	of New South Wales, Sydney, Australia. His current
	research interests include UAV communications and
	5G and beyond. 
	
	Dr. Zhang is the Vice President of the IEEE Communications Society. He has received six best paper
	awards from IEEE conferences and ComSoc technical committees. Within the IEEE ComSoc, he has taken many leadership positions, including a Member-at-Large on the Board of Governors (2018–2020),
	the Chair of the Wireless Communications Technical Committee (2019–2020), the Vice Director of Asia–Pacific Board (2016–2021), the Editor-in-Chief of
	IEEE WIRELESS COMMUNICATIONS LETTERS (2016–2019), the Technical Program Committee Chair of APCC 2017 and ICCC 2019, the Award Committee Chair of Asia–Pacific Board, and the Award Committee Chair of Technical Committee on Cognitive Networks. In addition, he has served
	as a member for various ComSoc boards/standing committees, including Journals Board, Technical Committee Recertification Committee, Finance Standing Committee, Information Technology Committee, Steering Committee of IEEE TRANSACTIONS ON GREEN COMMUNICATIONS AND NETWORKING, and Steering Committee of IEEE NETWORKING LETTERS.
	He serves as an Area Editor for the IEEE TRANSACTIONS ON WIRELESS COMMUNICATIONS and the Editor-in-Chief for Journal of Communications and Information Networks. Previously, he has served as an Editor for IEEE TRANSACTIONS ON COMMUNICATIONS, IEEE TRANSACTIONS
	ON WIRELESS COMMUNICATIONS, IEEE TRANSACTIONS ON COGNITIVE
	COMMUNICATIONS AND NETWORKING, and IEEE JOURNAL ON SELECTED
	AREAS IN COMMUNICATIONS—Cognitive Radio Series. He was an IEEE ComSoc Distinguished Lecturer in 2016–2017.
\end{IEEEbiography}

\end{document}